\documentclass[preprint2]{./proto}
\usepackage{times}

\newcommand{\refs}{\par\noindent\hangindent=1pc\hangafter=1}
\begin{document}

%\title{\textbf{\LARGE Spatially Resolved Observations of the Close 
%Circumstellar Environment of Young Stars and Implications for Star, Disk and 
%Planet Formation}}

\title{\textbf{\LARGE The Circumstellar Environments of Young Stars at AU Scales}}

\author {\textbf{\large Rafael Millan-Gabet}}
\affil{\small\em California Institute of Technology}

\author {\textbf{\large Fabien Malbet}}
\affil{\small\em Laboratoire d'Astrophysique de Grenoble}

\author {\textbf{\large Rachel Akeson}}
\affil{\small\em California Institute of Technology}

\author {\textbf{\large Christoph Leinert}}
\affil{\small\em Max-Planck-Institut f\"{u}r Astronomie}

\author {\textbf{\large John Monnier}}
\affil{\small\em University of Michigan}

\author {\textbf{\large Rens Waters}}
\affil{\small\em University of Amsterdam}

%\author {\textbf{\large DRAFT February 23 2006}}

\begin{abstract}
%\begin{list}{ } {\rightmargin 1in}
%{\leftmargin 0in}
\baselineskip = 11pt
%rule{4.75in}{0.5pt}
%\vskip 1pt
\leftskip = 0.65in
\rightskip = 0.65in \parindent=1pc {\small We review recent advances
in our understanding of the innermost regions of the circumstellar
environment around young stars, made possible by the technique of long
baseline interferometry at infrared wavelengths. Near-infrared
observations directly probe the location of the hottest dust. The
characteristic sizes found are much larger than previously thought,
and strongly correlate with the luminosity of the central young
stars. This relation has motivated in part a new class of models of
the inner disk structure. The first mid-infrared observations have
probed disk emission over a larger range of scales, and spectrally
resolved interferometry has for the first time revealed mineralogy
gradients in the disk.  These new measurements provide crucial
information on the structure and physical properties of young
circumstellar disks, as initial conditions for planet formation.
%In addition to summarizing these pioneering observations, we expose the
%many open questions that accompany the impressive progress made, and
%anticipate the experimental and modelling efforts that promise to help
%elucidate the diverse phenomena associated with the close
%circumstellar environment of young stars.
\\~\\~\\~}%leave this in to get the correct vertical space after the abstract

%\end{list}
\end{abstract}  

\section{\textbf{INTRODUCTION}}

Stars form from collapsing clouds of gas and dust and in their
earliest infancies are surrounded by complex environments that obscure
our view at optical wavelengths. As evolution proceeds, a stage is
revealed with three main components: the young star, a circumstellar
disk and an infalling envelope. Eventually, the envelope dissipates,
and the emission is dominated by the young star-disk system. Later on,
the disk also dissipates to very tenuous levels.  It is out of the
young circumstellar disks that planets are expected to form, and
therefore understanding their physical conditions is necessary before
we can understand the formation process.  Of particular interest are
the inner few AU (Astronomical Unit), corresponding to formation sites
of Terrestrial type planets, and to migration sites for gas giants
presumably formed further out in the disk (see the chapter by {\em
Udry, Fisher and Queloz}).

A great deal of direct observational support exists for the scenario
outlined above, and in particular for the existence of circumstellar
disks around young stars (see the chapters by {\em Dutrey et al.} and
by {\em Watson et al.}). In addition, the spectroscopic and
spectro-photometric characteristics of these systems (i.e. spectral
energy distributions, emission lines) are also well described by the
disk hypothesis. However, current optical and millimeter-wave imaging
typically probe scales of 100$-$1000s of AU with resolutions of 10s of
AU, and models of unresolved observations are degenerate with respect
to the spatial distribution of material. As a result, our
understanding of even the most general properties of the circumstellar
environment at few~AU or smaller spatial scales is in its infancy.

\begin{figure}[htbp]
\epsscale{1.}
\plotone{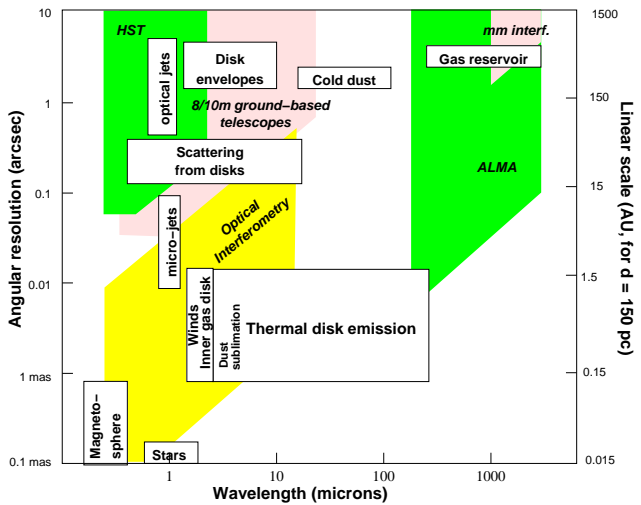}
\caption{\small
Observational phase-space (spectral domain and angular resolution) for
optical interferometers, and for complementary techniques (shaded
polygons). Also outlined over the most relevant phase-space regions
(rectangular boxes) are the main physical phenomena associated with
young stellar objects.
\label{phase_space_fig}}  
\end{figure}

Currently, the only way to achieve sufficient angular resolution to
directly reveal emission within the inner AU is through interferometry
at visible and infrared wavelengths, here referred to as {\em optical}
interferometry.  An interferometer with a baseline length of $B=100$~m
(typical of current facilities) operating at near to mid-infrared
wavelengths (NIR, MIR, typically H to N bands, $\lambda_0 = 1.65 - 10
\, \mu m$) probes $1800 - 300$~K material and achieves an angular
resolution $\sim \lambda_0/2B \sim 2 - 10$~milliarcseconds (mas), or
$0.25-1.5$~AU at a distance typical of the nearest well known star
forming regions (150~pc).  This observational discovery space is
illustrated in Fig.~\ref{phase_space_fig}, along with the domains
corresponding to various young stellar object (YSO) phenomena and
complementary techniques and instruments. Optical interferometers are
ideally suited to directly probe the innermost regions of the
circumstellar environment around young stars, and indeed using this
technique surprising and rapid progress has been made, as the results
reviewed in this chapter will show.

Optical, as well as radio, interferometers operate by coherently
combining the electromagnetic waves collected by two or more
telescopes. Under conditions that apply to most astrophysical
observations, the amplitude and phase of the resulting interference
patterns (components of the complex {\em visibility}) are related via
a two-dimensional Fourier transform to the brightness distribution of
the object being observed. For a detailed description of the
fundamental principles, variations in their practical implementation,
and science highlights in a variety of astrophysics areas we refer to
the reviews by {\em Monnier}~(2003) and {\em Quirrenbach}~(2001). The
reader may also be interested in consulting the proceedings of topical
summer schools and workshops such as {\em Lawson}~(2000), {\em Perrin
and Malbet}~(2003), {\em Garcia et al.}~(2003) and {\em Paresce et
al.}~(2006) (online proceedings of the Michelson Summer Workshops may
also be found at {\tt http://msc.caltech.edu/michelson/}).

While interferometers are capable of model-independent imaging by
combining the light from many telescopes, most results to date have
been obtained using two-telescopes (a single-baseline). The main
characteristics of current facilities involved in these studies are
summarized in Table~\ref{tbl-1}.  Interferometer data, even with
sparse spatial frequency coverage, provides direct constraints about
source geometry and thus contains some of the power of direct
imaging. However, with such data alone only the simplest objects can
be constrained.  Therefore, more typically, a small number of fringe
visibility amplitude data points are combined with a spectral energy
distribution (SED) for fitting simple physically-motivated geometrical
models; such as a point source representing the central star
surrounded by a Gaussian or ring-like brightness (depending on the
context) representing the source of infrared excess flux.  This allows
us to determine ``characteristic sizes" at the wavelength of
observation for many types of YSOs (e.g., T~Tauri, Herbig~Ae/Be,
FU~Orionis).  The interferometer data can also can be compared to
predictions of specific physical models and this exercise has allowed
to rule out certain classes of models, and provided crucial
constraints to models that can be made to reproduce the spectral and
spatial observables.

A note on nomenclature: Before the innermost disk regions could be
spatially resolved, as described in this review, disk models were
tailored to fit primarily the unresolved spectro-photometry.  The
models used can be generally characterized as being geometrically thin
and optically thick, with simple radial temperature power laws with
exponents $q = -0.75$ (flat disk heated by accretion and/or stellar
irradiation, {\em Lynden-Bell and Pringle},~1974) or shallower (flared
disks, {\em Kenyon and Hartmann},~1987; {\em Chiang and
Goldreich},~1997). In this review, we will refer to these classes of
models as ``classical'', to be distinguished from models with a
modified inner disk structure that have recently emerged in part to
explain the NIR interferometer measurements.

The outline of this review is as follows. Sections~2~and~3 summarize
the observational state of the art and emerging interpretations,
drawing a physically motivated distinction between inner and outer
disk regions that also naturally addresses distinct wavelength regimes
and experimental techniques. We note that these developments are
indeed recent, with the first preliminary analyses of interferometric
observations of YSOs being presented (as posters) at the Protostars
and Planets IV Conference in 1998. In Section~4 we highlight
additional YSO phenomena that current facilities are addressing, such
as the disk-wind connection and young star mass determination, using
combination of interferometry with high resolution line spectroscopy.
In Section~5 we present a top-level summary of the main
well-established results, and highlight remaining important open
questions likely to be experimentally and theoretically addressed in
the coming years.  In Section~6 we describe additional phenomena yet
to be explored and new instrumental capabilities that promise to
enable this future progress.

%%%%%%%%%%%%%%%%%%%%%%%%%%%%%%%%%%%%%%%%%%%%%%%%%%%%%%%%%%%%%%%%%%%%%%%

%%%%%%%%%%%%%%%%%%%%%%%%%%%%%%%%%%%%%%%%%%%%%%%%%%%%%%%%%%%%%%%%%%%%%%%

\begin{deluxetable}{lllcccc}
\tabletypesize{\small}
\tablecaption{Long Baseline Optical Interferometers Involved in YSO Research\label{tbl-1}}
\tablewidth{0pt}
\tablehead{Facility & Instrument\tablenotemark{a} & Wavelength & Number of & 
Telescope & Baseline Range & Best Resolution\tablenotemark{d}\\ 
&  & Coverage\tablenotemark{b}  & Telescopes\tablenotemark{c} & Diameter (m) & (m) & (mas)}
\startdata

PTI  & $V^2$         & $1.6-2.2 \mu m$ [44]      & 2 [3]   & 0.4       & 80 $-$ 110 & 1.5  \\
IOTA & $V^2$, IONIC3 & $1.6-2.2 \mu m$           & 3       & 0.4       & 5  $-$ 38  & 4.5  \\
ISI  & Heterodyne    & $11 \mu m$                & 3       & 1.65      & 4 $-$ 70   & 16.2 \\
KI   & $V^2$         & $1.6-2.2 \mu m$ [22]      & 2       & 10        & 85         & 2.0  \\
KI   & Nuller        & $8-13 \mu m$    [34]      & 2       & 10        & 85         & 9.7 \\
VLTI & MIDI          & $8-13 \mu m$    [250]     & 2 [8]   & 8.2 / 1.8 & 8 $-$ 200  & 4.1  \\
VLTI & AMBER         & $1-2.5 \mu m$   [$10^4$]  & 3 [8]   & 8.2 / 1.8 & 8 $-$ 200  & 0.6  \\
CHARA & $V^2$        & $1.6-2.2 \mu m$           & 2 [6]   & 1         & 50 $-$ 350 & 0.4  \\

\enddata
\tablecomments{
{\bf (a)}
$\, V^2$ refers to a mode in which only the visibility amplitude is measured, often implemented
in practice as a measurement of the (un-biased estimator) square of the visibility amplitude.
{\bf (b)}
The maximum spectral resolution available is given in square brackets.
{\bf (c)}
The number of telescopes that can be simultaneously combined is given, along with
the total number of telescopes available in the array in square brackets.
{\bf (d)}
In each case, the best resolution is given as half the fringe spacing, 
$\lambda_{min}/(2 B_{max})$, for the longest physical baseline length and shortest 
wavelength available.
{\bf References:}
[1] PTI, {\em Palomar Testbed Interferometer, Colavita et al}~1999.
[2] IOTA, {\em Infrared-Optical Telescope Array, Traub et al.}~2004. 
[3] ISI, {\em Infrared Spatial Interferometer, Hale et al.}~2000.
[4] KI, {\em Keck Interferometer, Colavita, Wizinovich and Akeson}~2004.
[5] KI Nuller, {\em Serabyn et al.}~2004.
[6] MIDI, {\em MID-infrared Interferometric instrument, Leinert et al.}~2003.
[7] AMBER, {\em Astronomical Multiple BEam Recombiner, Malbet et al.}~2004.
[8] CHARA, {\em Center for High Angular Resolution Astronomy interferometer, ten Brummelaar et al.}~2005.
}

%\tablerefs{
% [1] PTI, {\em Palomar Testbed Interferometer, Colavita et al}~1999.
% [2] IOTA, {\em Infrared-Optical Telescope Array, Traub et al.}~2004. 
% [3] ISI, {\em Infrared Spatial Interferometer, Hale et al.}~2000.
% [4] KI, {\em Keck Interferometer, Colavita, Wizinovich and Akeson}~2004.
% [5] KI Nuller, {\em Serabyn et al.}~2004.
% [6] MIDI, {\em MID-infrared Interferometric instrument, Leinert et al.}~2003.
% [7] AMBER, {\em Astronomical Multiple BEam Recombiner, Malbet et al.}~2004.
% [8] CHARA, {\em Center for High Angular Resolution Astronomy interferometer, ten Brummelaar et al.}~2005.
% [9] MIRC, {\em Michigan Infrared Combiner, Monnier et al.}~2004.
% }
\end{deluxetable}

%\tablenotetext{a}{
%The maximum spectral resolution available is given in parenthesis}
%\tablenotetext{b}{
%The number of telescopes that can be simultaneously combined is given, along with
%the total number of telescopes available in the array in parenthesis.}
%\tablenotetext{c}{
%In each case, the best resolution is given as half the fringe spacing, $\lambda_{min}/(2 B_{max})$, for the
%longest physical baseline length and shortest wavelength available.}

%%%%%%%%%%%%%%%%%%%%%%%%%%%%%%%%%%%%%%%%%%%%%%%%%%%%%%%%%%%%%%%%%%%%%%%

\bigskip
\centerline{\textbf{ 2. THE INNER DISK}}
\bigskip

At NIR wavelengths ($\sim 1 - 2.4 \, \mu m$) optical interferometers
probe the hottest circumstellar material located at stellocentric
distances $\lesssim 1$~AU.  In this review these regions are referred
to as the {\it inner disk}. For technical reasons, it is at NIR
wavelengths that interferometric measurements of YSOs first became
possible, and provided the first direct probes of inner disk
properties.  Until only a few years ago, relatively simple models of
accretion disks were adequate to reproduce most observables. However,
the new observations with higher spatial resolution have revealed a
richer set of phenomena.

\bigskip
\noindent
\textbf{ 2.1 Herbig Ae/Be Objects}
\bigskip

% maybe missing a comment on how the observational work had no prior
% theoretical guidance.

In Herbig Ae/Be (HAeBe) objects, circumstellar material
(proto-planetary disk, remnant envelope, or both) surrounds a young
star of intermediate mass ($\sim 2-10 \, M_{\sun}$, see e.g., the review
by {\em Natta et al.}~2000). A significant number of
these objects fall within the sensitivity limits of small
interferometers, hence the first YSO studies containing relatively
large samples focused on objects in this class.

AB~Aur was the first ``normal'' (i.e. non-outburst, see Section~2.3)
YSO to be resolved at NIR wavelengths.  In observations at the {\em
Infrared Optical Telescope Array} (IOTA) AB~Aur appeared clearly
resolved in the H and K spectral bands on 38~m baselines, by an amount
corresponding to a characteristic diameter of 0.6~AU for the
circumstellar NIR emission ({\em Millan-Gabet et al.},~1999).  The
measured size was unexpectedly large in the context of then-current
disk models of HAeBe objects (e.g., {\em Hillenbrand et al.},~1992)
which predicted NIR diameter of 0.2~AU based on optically-thick,
geometrically-thin circumstellar disk with a small dust-free inner
hole.  This conclusion was reinforced by the completed IOTA survey of
{\em Millan-Gabet et al.}~(2001), finding characteristic NIR sizes of
0.6$-$6~AU for objects with a range of stellar properties (spectral
types A2--O9).

Due to the limited position angle coverage of these first
single-baseline observations, the precise geometry of the NIR emission
remained ambiguous. {\em Millan-Gabet et al.}~(2001) found that the
NIR sizes were similar at 1.65 and $2.2 \, \mu m$, suggesting a steep
temperature gradient at the inner disk edge.  These early observations
revealed no direct evidence for ``disk-like" morphologies; indeed, for
the few objects with size measurements at multiple position angles,
the data indicated circular symmetry (as if from spherical halos or
face-on disks, but not inclined disks).

The first unambiguous indication that the NIR emission arises in
disk-like structures came from single-telescope imaging of the
highly-luminous YSOs LkH$\alpha$~101 and MWC~349-A, using a
single-telescope interferometric technique (aperture masking at the
Keck~I telescope).  MWC~349-A appeared clearly elongated ({\em Danchi
et al.},~2001) and LkH$\alpha$~101 presented an asymmetric ring-like
morphology, interpreted as a bright inner disk edge occulted on one
side by foreground cooler material in the outer regions of the flared
disk ({\em Tuthill et al.},~2001). Moreover, the location of the
LkH$\alpha$~101 ring was also found to be inconsistent with
predictions of classical disk models and these authors suggested that
an optically-thin inner cavity (instead of optically-thick disk
midplane) would result in larger dust sublimation radii, consistent
with the new size measurements.

Could simple dust sublimation of directly heated dust be setting the
NIR sizes of other HAeBe objects as well?  The idea was put to the
test by {\em Monnier and Millan-Gabet}~(2002). Inspired by the
LkH$\alpha$~101 morphology and interpretation, these authors fit a
simple model consisting of a central star surrounded by a thin ring
for all objects that were measured interferometrically at the time
(IOTA and aperture masking, plus first YSO results from the {\em
Palomar Testbed Interferometer}, PTI, {\em Akeson et
al.},~2000). Indeed, the fitted ring radii are clearly correlated with
the luminosity of the central star, and follow the expected relation
$R \propto L_{\star}^{1/2}$. Further, by using realistic dust
properties it was also found that the measured NIR sizes are
consistent with the dust sublimation radii of relatively large grains
($\gtrsim 1 \, \mu m$) with sublimation temperatures in the range
1000~K$-$2000~K.

\begin{figure*}[htbp]
\epsscale{2.0}
\plotone{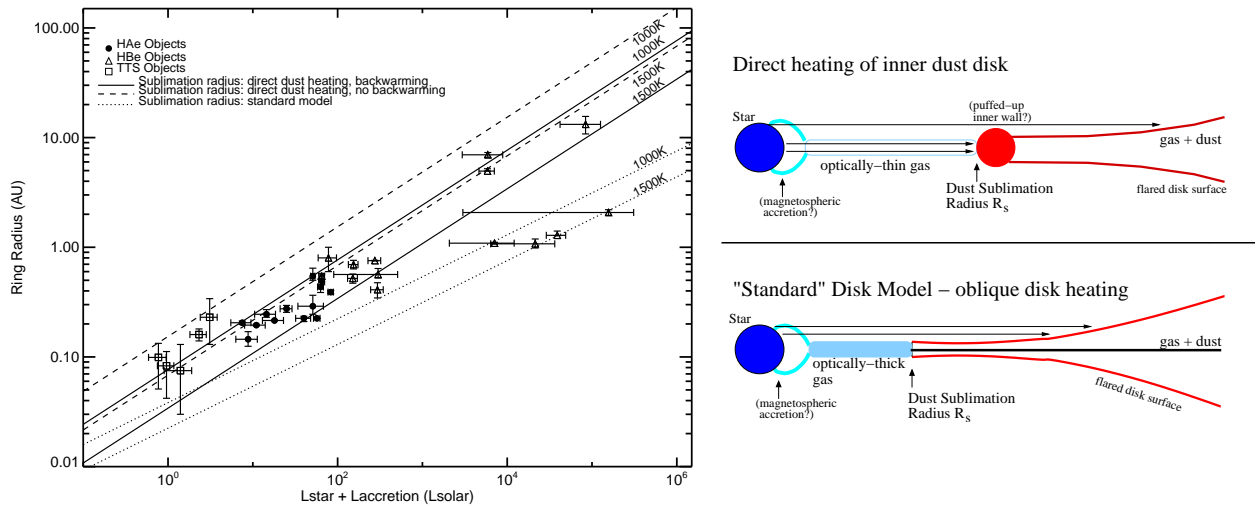}
\caption{\small Measured sizes of HAeBe and T~Tauri objects {\em vs.}
central luminosity (stellar $+$ accretion shock); and comparison with
sublimation radii for dust directly heated by the central luminosity
(solid and dashed lines) and for the oblique heating implied by the classical
models (dotted line).  A schematic representation of the key features
of inner disk structure in these two classes of models is also
shown. Data for HAeBe objects are from: IOTA ({\em Millan-Gabet et
al.},~2001), Keck aperture masking ({\em Danchi et al.},~2001, {\em
Tuthill et al.},~2001), PTI ({\em Eisner et al.},~2004) and KI ({\em
Monnier et al.},~2005a).  Data for T~Tauri objects are from PTI ({\em
Akeson et al.},~2005b) and KI ({\em Akeson et al.},~2005a, {\em Eisner et
al.},~2005). For clarity, for objects observed at more than one
facility, we include only the most recent measurement.
\label{size-fig}}  
\end{figure*}

Following {\em Monnier and Millan-Gabet}~(2002), in
Fig.~\ref{size-fig} we have constructed an updated diagram of NIR size
{\em vs.} central luminosity based on all existing data in the
literature. We include data for both HAeBe and T~Tauri objects, the
latter being discussed in detail in the next Section; and we also
illustrate schematically the essential ingredients (i.e. location of
the inner dust disk edge) of the models to which the data are being
compared.  
%For HAeBe objects, in addition to the references above, we
%include additional measurements from PTI ({\em Eisner et al.},~2003,
%2004) and from recent observations using the {\em Keck Interferometer}
% (KI) ({\em Monnier et al.},~2005). 
For HAeBe objects, disk irradiation
is dominated by the stellar luminosity, and therefore we neglect
heating by accretion shock luminosity (which may play a significant
role for the most active lower stellar luminosity T~Tauri objects, as
discussed in the next section).

Indeed, it can be seen that for HAe and late HBe objects, over more
than two decades in stellar luminosity, the measured NIR sizes are
tightly contained within the sublimation radii of directly heated grey
dust with sublimation temperatures of 1000~K$-$1500~K under the
assumption that dust grains radiate over the full solid angle (e.g.,
no back-warming, solid lines in Fig.~\ref{size-fig}).  If instead we
assume that the dust grains emit only over 2$\pi$ steradian (e.g.,
full back-warming) then the corresponding sublimation temperature
range is 1500~K$-$2000~K.

The most luminous objects (the early spectral type HBe objects) on the
other hand, have NIR sizes in good agreement with the classical model,
indicating that for these objects the disk extends closer in to the
central star.  {\em Monnier and Millan-Gabet}~(2002) hypothesized that
even optically thin low-density gas inside the dust destruction radius
can scatter UV stellar photons partly shielding the dust at the inner
disk edge and reducing the sublimation radius. {\em Eisner et
al.}~(2003, 2004) argue instead that the distinction may mark a
transition from magnetospheric accretion (late types) to a regime for
the early types where optically-thick disks in fact do extend to the
stellar surface, either as a result of higher accretion rates or weak
stellar magnetic fields.  We note that using polarimetry across
H$_\alpha$ lines originating in the inner gas disk at $\sim$ few
$R_\star$, {\em Vink et al.}~(2005) (and references therein) also find
(in addition to flattened structures consistent with disk-like
morphology) a marked difference in properties between HBe objects and
HAe and T~Tauri objects; qualitatively consistent with the NIR size
properties described above.
%consistent with the presence presence of an
%un-disrupted gas disk on small scales for HBe objects, while the less
%massive objects (HAe and T~Tauri) show evidence for line photons
%scattering off an external rotating disk.
However, not {\em all} high luminosity HBe objects are in better
agreement with the classical model; there are notable exceptions
(e.g., LkH$\alpha$~101, MWC~349-A, see Fig.~\ref{size-fig}) and {\em
Monnier et al.}~(2005a) have noted that these may be more evolved
systems, where other physical processes, such as dispersal by stellar
winds or erosion by photo-evaporation, have a dominant effect in
shaping the inner disk.

In parallel with these developments, and in the context of detailed
modelling of the NIR SED, {\em Natta et al.}~(2001) similarly
formulated the hypothesis that the inner regions of circumstellar
disks may be largely optically thin to the stellar photons such that a
directly illuminated ``wall'' forms at the inner dust-disk edge. The
relatively large surface area of this inner dust wall is able to
produce the required levels of NIR flux (the so-called NIR SED
``bump''). The simple blackbody surface of the initial modelling
implementation has since been replaced by more realistic models which
take additional physical effects into account: self-consistent
treatment of radiative transfer and hydrostatic vertical structure of
the inner wall and corresponding outer disk shadowing ({\em Dullemond
et al.},~2001), optical depth and scale height of gas inside the dust
destruction radius ({\em Muzerolle et al.},~2004), and inner rim
curvature due to density-dependent dust sublimation temperature ({\em
Isella and Natta},~2005).

Independent of model details however, it is clear that the
optically-thin cavity disk model with a ``puffed-up'' inner dust wall
can explain a variety of observables: the ring-like morphology, the
characteristic NIR sizes and the NIR SED. We also emphasize that
although the discussion of the interferometer data presented above was
centered around the size-luminosity diagram, the same conclusions have
been obtained by modelling SEDs and visibility measurements for
individual sources using physical models of both classical and
puffed-up inner rim models.

Although the characteristic NIR sizes have now been well established
for a relatively large sample, few constraints still exist as to the
actual geometry of the resolved NIR emission. In this respect, most
notable are the PTI observations of {\em Eisner et al.}~(2003, 2004)
which provided the first evidence for elongated emission, consistent
with inclined disks.  Further, the inclinations found can explain the
residual scatter (50\%) in the size-luminosity relation for HAe and
late HBe systems ({\em Monnier et al.},~2005a).

It is important to note that three different interferometers have now
reached the same conclusions regarding the characteristic NIR sizes of
HAeBe objects, in spite of having vastly different field-of-views
(FOV; 3, 1, and 0.05~{\em arcsec} FWHM for IOTA, PTI and KI
respectively). This indicates that the interpretation is not
significantly effected by non-modelled flux entering the FOV, from the
large scale (scattered) emission often found around these objects
(e.g., {\em Leinert et al.},~2001). Indeed, if not taken into account
in a composite model, such incoherent flux would lead to
over-estimated characteristic sizes, and still needs to be considered
as a (small) possible effect for some objects.

\bigskip
\noindent
\textbf{ 2.2 T Tauri Objects}
\bigskip

Only the brightest T~Tauri objects can be observed with the small
aperture interferometers, so early observations were limited in
number.  The first observations were taken at the PTI and were of the
luminous sources T~Tau~N and SU~Aur ({\em Akeson et al.},~2000,
2002). Later observations ({\em Akeson et al.},~2005a) added DR~Tau and
RY~Tau to the sample, and employed all three PTI baselines to
constrain the disk inclination. Using simple geometrical models (or
simple power-law disk models) the characteristic NIR sizes (or disk
inner radii) were found to be larger than predicted by SED-based
models ({\em e.g., Malbet and Bertout},~1995), similar to the result
for the HAe objects.  For example, inclined ring models for SU~Aur and
RY~Tau yield radii of 10~R$_{\star}$ and 11~R$_{\star}$ respectively,
substantially larger than magnetic truncation radii of
3--5~R$_{\star}$ expected from magnetospheric accretion models ({\em
Shu et al.},~1994).  If the inner dust disk does not terminate at the
magnetic radius, what then sets its radial location?  Interestingly,
these initial T~Tauri observations revealed NIR sizes that were also
consistent with sublimation radii of dust directly heated by the
central star, suggesting perhaps that similar physical processes,
decoupled from magnetospheric accretion, set the NIR sizes of both
T~Tauri and HAe objects.  {\em Lachaume et al.}~(2003) were the first
to use physical models to fit the new visibility data, as well as the
SEDs. Using a two-layer accretion disk model, these authors find
satisfactory fits for SU~Aur (and FU~Ori, see Section~2.3), in
solutions that are characterized by the mid-plane temperature being
dominated by accretion, while the emerging flux is dominated by
reprocessed stellar photons.

Analysis of T~Tauri systems of more typical luminosities became
possible with the advent of large aperture infrared interferometers.
Observations at the KI ({\em Colavita et al.},~2003; {\em Eisner et
al.},~2005; {\em Akeson et al.},~2005a) continued to find large NIR
sizes for lower luminosity stars, in many cases even larger than would
be expected from extrapolation of the HAe relation.

Current T~Tauri measurements and comparison with dust sublimation
radii are also summarized in Fig.~\ref{size-fig}.  {\em Eisner et
al.}~(2005) found better agreement with puffed-up inner wall models,
and in Fig.~\ref{size-fig} we use their derived inner disk radii.
Following {\em Muzerolle et al.}~(2003) and {\em D'Alessio et
al.}~(2004), the central luminosity for T~Tauri objects includes the
accretion luminosity, released when the magnetic accretion flow
encounters the stellar surface, which can contribute to the location
of the dust sublimation radius for objects experiencing the highest
accretion rates (in Fig.~\ref{size-fig}, accretion luminosity only
affects the location of three objects significantly: AS205--A, PX~Vul
and DR~Tau, having L$_{accretion}$/L$_{\star} \sim 10.0, 1.8, 1.2$
respectively.)

It can be seen that many T~Tauri objects follow the HAe relation,
although several are even larger given the central luminosity (the
T~Tauri objects located above the 1000~K solid line in
Fig.~\ref{size-fig} are: BP~Tau, GM~Aur, LkCa~15 and RW~Aur).

It must be noted however that current measurements of the lowest
luminosity T~Tauri objects have relatively large errors.  Further, the
ring radii plotted in Fig.~\ref{size-fig} depend on the relative
star/disk NIR flux contributions, most often derived from an SED
decomposition which for T~Tauri objects is considerably more uncertain
than for the HAeBe objects, due to their higher photometric
variability, and the fact that the stellar SEDs peak not far ($\sim 1
\, \mu m$) from the NIR wavelengths of observation. We note that an
improved approach was taken by {\em Eisner et al.}~(2005), who derived
stellar fluxes at the KI wavelength of observation from near
contemporaneous optical veiling measurements and extrapolation to NIR
wavelengths based on model atmospheres. These caveats illustrate the
importance of obtaining more direct estimates of the excess NIR
fluxes, via veiling measurements (e.g., {\em Johns-Krull and
Valenti},~2001 and references therein).

Under the interpretation that the T~Tauri NIR sizes trace the dust
sublimation radii, different properties for the dust located at the
disk inner edge appear to be required to explain all objects.  In
particular, if the inner dust-disk edge becomes very abruptly
optically thick (e.g., strong backwarming), the sublimation radii
become larger by a factor of $\times 2$, in better agreement with some
of the low luminosity T~Tauri objects (dashed lines in
Fig.~\ref{size-fig}). As noted by {\em Monnier and
Millan-Gabet}~(2002), smaller grains also lead to larger sublimation
radii, by a factor $\sqrt{Q_R}$, where $Q_R$ is the ratio of dust
absorption efficiencies at the temperatures of the incident and
re-emitted field ($Q_R \simeq 1 - 10$ for grain sizes $1.0 - 0.03 \, \mu
m$, the incident field of a stellar temperature of 5000~K typical of
T~Tauri stars, and re-emission at 1500~K).

Alternatively, an evolutionary effect could be contributing to the
size-luminosity scatter for the T~Tauri sample. {\em Akeson et
al.}~(2005a) note that the size discrepancy (measured {\em vs.}
sublimation radii) is greatest for objects with the lowest ratio of
accretion to stellar luminosity. If objects with lower accretion rates
are older (e.g., {\em Hartmann et al.},~1998) then other physical
processes such as disk dispersal ({\em Clarke et al.},~2001, see also
the review by {\em Hollenbach et al.},~2000) may be at play for the
most ``discrepant'' systems mentioned above (the size derived for
GM~Aur contains additional uncertainties, given the evidence in this
system for a dust gap at several~AU, {\em Rice et al.},~2003, and the
possibility that it contributes significant scattered NIR light).

In detailed modelling work of their PTI data, {\em Akeson et
al.}~(2005b) used Monte Carlo radiative transfer calculations ({\em
Bjorkman and Wood},~2001) to model the SEDs and infrared visibilities
of SU~Aur, RY~Tau and DR~Tau. These models addressed two important
questions concerning the validity of the size interpretation based on
fits to simple star~$+$~ring models: (1) Is there significant NIR
thermal emission from {\em gas} inside the dust destruction radius?
and, (2) is there significant flux arising in large scale (scattered)
emission?  The modelling code includes accretion and shock/boundary
luminosity as well as multiple scattering and this technique naturally
accounts for the radiative transfer effects and the heating and
hydrostatic structure of the inner wall of the disk.  In these models,
the gas disk extends to within a few stellar radii and the dust disk
radius is set at the dust sublimation radius.  For SU~Aur and RY~Tau,
the gas within the inner dust disk was found to contribute
substantially to the NIR emission.  Modelling of HAeBe circumstellar
disks by {\em Muzerolle et al.}~(2004) found that emission from the
inner gas exceeded the stellar emission for accretion rates $>10^{-7}$
M$_{\odot}$ yr$^{-1}$. Thus, even as a simple geometrical
representation, a ring model at the dust sublimation temperature may
be too simplistic for sources where the accretion luminosity is a
substantial fraction of the total system luminosity.  Moreover, even
if the NIR gas emission were located close to the central star and
were essentially unresolved, its presence would affect the SED
decomposition (amount of NIR excess attributed to the dust ``ring''),
leading to incorrect (under-estimated) ring radii. The second finding
from the radiative models is that the extended emission (here meaning
emission from scales larger than 10~mas) was less than 6\% of the
total emission for all sources, implying that scattered light is not a
dominant component for these systems.

Clearly, the new spatially resolved observations just described,
although basic, constitute our only direct view into the physical
conditions in young disks on scales $\lesssim 1$~AU, and therefore
deliver crucial ingredients as pre-conditions for planet formation.
As pointed out by {\em Eisner et al.}~(2005) the interpretation that
the measured NIR sizes correspond to the location of the innermost
disk dust implies that dust is in fact present at Terrestrial planet
locations. Conversely, the measured radii imply that no dust exists,
and therefore planet formation is unlikely, inside these relatively
large $\sim$ 0.1--1~AU inner dust cavities. Finally, that the dust
disk stops at these radii may also have implications for planet
migration, depending on whether migration is halted at the inner
radius of the gas ({\em Lin et al.},~1996) or dust ({\em Kuchner and
Lecar},~2002) disk (see also the chapter by {\em Papaloizou et al.}).

%%%%%%%%%%%%%%%%%%%%%%%%%%%%%%%%%%%%%%%%%%%%%%%%%%%%%%%%%%%%%%%%%%%%%%%

% {\em Muzerolle et al.}~2003 have indeed proposed that the
%``puffed-up'' inner wall scenario also applies to T~Tauri objects,
%even if optically thick gas exists between the star and the inner dust
%edge ({\em Muzerolle et al.}~2004).  In this case however, accretion
%luminosity (released when the magnetic accretion flow encounters the
%stellar surface) can be a significant additional central heating
%source, contributing to the location of the dust sublimation radius
%for objects experiencing the highest accretion. Using this model, {\em
%Muzerolle et al.}~2003 and {\em D'Alessio}~2004 made detailed
%predictions concerning the location of the dust disk inner rim based
%on fits to SEDs, and comparison with existing measurements is
%encouraging. For some objects (DG~Tau, LkCa15, DR~Tau, BP~Tau) the
%model rim radii agree well (8--40\%) with ring radii derived from the
%interferometer data, although in other cases (SU~Aur, RY~Tau)
%predictions and measurements are very discrepant. This comparison is
%of course over-simplistic, and the situation may become more clear
%when these (and other) detailed models are fit directly to the
%interferometer visibilities in addition to the spectro-photometry.

%%%%%%%%%%%%%%%%%%%%%%%%%%%%%%%%%%%%%%%%%%%%%%%%%%%%%%%%%%%%%%%%%%%%%%%

%%%%%%%%%%%%%%%%%%%%%%%%%%%%%%%%%%%%%%%%%%%%%%%%%%%%%%%%%%%%%%%%%%%%%%%

\bigskip
\noindent
\textbf{2.3 Detailed Tests of the Disk Hypothesis: FU Orionis}
\bigskip

FU~Orionis objects are a rare type of YSO believed to be T~Tauri stars
surrounded by a disk that has recently undergone an episode of
accretion rate outburst (see e.g., the review by {\em Hartmann et
al.},~1996). During the outburst, the system brightens by several
visual magnitudes, followed by decade-long fading.  The disk
luminosity dominates the emission at all wavelengths, and is expected
to be well represented by thermal emission by disk annulii that follow
the canonical temperature profile $T \propto r^{-3/4}$. With this
prescription, the model has few parameters that are well constrained
by the SEDs alone (except for inclination effects).  In principle
then, these systems are ideal laboratories for testing the validity of
disk models. The total number of known objects in this class is small
(e.g., 20 objects in the recent compilation by {\em \'{A}brah\'{a}m et
al.},~2004), and the subset that is observable by current optical
interferometers is even smaller (about 6 objects, typically limited by
instrumental sensitivity at short wavelengths $0.55-1.25 \, \mu m$).

The very first YSO to be observed with an optical interferometer was
in fact the prototype for this class, FU~Orionis itself ({\em Malbet
et al.},~1998). Indeed, the first K-band visibility amplitudes
measured by the PTI agreed well with the predictions of the canonical
disk model.  This basic conclusion has been re-inforced in a more
recent study using multi-baseline, multi-interferometer observations
({\em Malbet et al.},~2005). The detailed work on FU~Ori will
ultimately close the debate on its nature: current evidence favors
that it is a T~Tauri star surrounded by a disk undergoing massive
accretion ({\em Hartmann and Kenyon},~1985, 1996) rather than a
rotating supergiant with extreme chromosphere activity ({\em Herbig et
al.},~2003); and under that interpretation the physical (temperature
law) and geometrical (inclination and orientation) parameters
describing the disk have been established with unprecedented detail.

Three more FU~Orionis objects have been resolved in the NIR: V1057~Cyg
(PTI, {\em Wilkin and Akeson},~2003; KI {\em Millan-Gabet et
al.},~2006), V1515~Cyg and ZCMa-SE ({\em Millan-Gabet et
al.},~2006). In contrast to the FU~Ori case, these three objects
appear more resolved in the NIR than expected from a $T \propto
r^{-3/4}$ model, and simple exponent adjustments of such a single
power-law model does not allow simultaneous fitting of the
visibilities and SEDs. On the other hand, these objects are also known
to possess large mid-infrared fluxes in excess of emission by flat
disks, usually attributed to a large degree of outer disk flaring or,
more likely, the presence of a dense dust envelope ({\em Kenyon and
Hartmann},~1991). The low visibilities measured, particularly for
V1057~Cyg and V1515~Cyg, can be explained by K-band flux ($\sim 10$\%
of total) in addition to the thermal disk emission, resulting from
scattering through envelope material over scales corresponding to the
interferometer FOV (50~mas FWHM). Contrary to initial expectations,
the likely complexity of the circumstellar environment of most
FU~Orionis objects will require observations at multiple baselines and
wavelengths in order to disentangle the putative multiple components,
discriminate between the competing models, and perform detailed tests
of accretion disk structure.

\bigskip
\centerline{\textbf{ 3. THE OUTER DISK}}
\bigskip

% midi gets V(lambda) at eacu uv!

\begin{figure}[h]
\epsscale{1.1}
%\plotone{./millan-gabet_fig3.ps}
\plotone{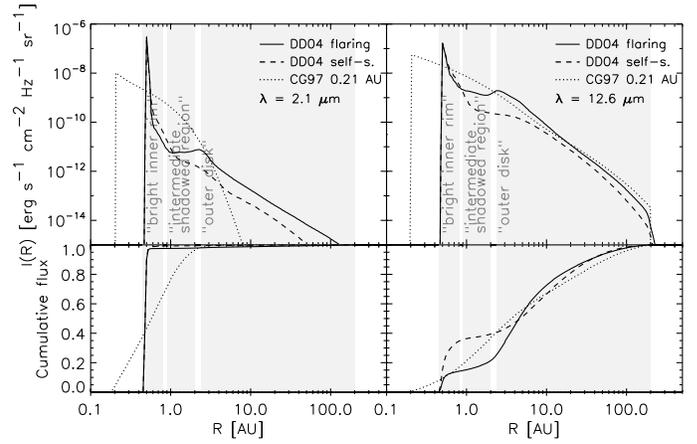}
\caption{\small 
Predicted radial distribution of NIR (left) and MIR (right) light for
typical disk models.  DD04 ({\em Dullemond and Dominik},~2004) refers
to a 2D radiative transfer and hydrostatic equilibrium Monte Carlo
code featuring a puffed up inner rim; CG97 to the gradually flaring
model of {\em Chiang and Goldreich}~1997.  We note that the CG97 disk
model can not self-consistently treat the inner disk once the
temperature of the ``super-heated'' dust layer increases beyond the dust
sublimation temperature. This only effects the NIR light profiles and
here the disk was artificially truncated at 0.21~AU.  The central star
is assumed to be of type A0 with M=2.5~M$_{\sun}$ and R = 2.0
R$_{\sun}$. The upper panels show the intensity profiles, and the
lower panels the cumulative brightness contributions, normalized to
1. Three qualitatively different disk regions are indicated by
shadowing.  This figure illustrates that NIR observations almost
exclusively probe the hot inner rim of the circumstellar disks, while
MIR observations are sensitive to disk structure over the first dozen
or so AU.  (Credit: Roy van Boekel).
\label{radial-fig}}
\end{figure}

At mid-infrared (MIR) wavelengths, and for 50$-$100~m baselines,
optical interferometers probe the spatial distribution and composition
of $\sim$ few 100~K circumstellar gas and dust, with $8-27$~mas
resolution, or $1-4$~AU at a distance of 150~pc.  The MIR radiation of
circumstellar disks in the N-band ($8-13 \, \mu m$) comes from a
comparatively wide range of distances from the star, see
Fig.~\ref{radial-fig}, and we now focus the discussion on
circumstellar dust in this several-AU transition region between the
hottest dust in the sublimation region close to the star (Section~2),
and the much cooler disk regions probed by mm~interferometry (see the
chapter by {\em Dutrey et al.}) and by scattered light imaging (see the
chapter by {\em Watson et al.}). As a matter of interpretational
convenience, at these longer MIR wavelengths the flux contribution
from scattered light originating from these relatively small spatial
scales is negligible, allowing to consider fewer model components and
physical processes.

\bigskip
\noindent
\textbf{ 3.1 Disk Sizes and Structure}
\bigskip

The geometry of proto-planetary disks is of great importance for a
better understanding of the processes of disk evolution, dissipation
and planet formation. The composition of the dust in the upper disk
layers plays a crucial role in this respect, since the optical and UV
photons of the central star, responsible for determining the local
disk temperature and thus the disk vertical scale-height, are absorbed
by these upper disk layer dust particles. The disk scale-height plays
a pivotal role in the process of dust aggregation and settling, and
therefore impacts the global disk evolution. In addition, the
formation of larger bodies in the disk can lead to structure, such as
density waves and gaps. 
%Clearly, high spatial resolution MIR imaging will be essential in
%order to probe these signs of planet formation.

Spatially resolved observations of disks in the MIR using single
telescopes have been limited to probing relatively large scale
(100s~AU) thermal emission in evolved disks around main-sequence stars
(e.g., {\em Koerner et al.},~1998; {\em Jayawardhana},~1998) or
scattering by younger systems (e.g., {\em McCabe et
al.},~2003). Single-aperture interferometric techniques (e.g.,
nulling, {\em Hinz et al.},~2001; {\em Liu et al.},~2003,~2005 or
segment-tilting, {\em Monnier et al.},~2004b) as well as observations
at the {\em Infrared Spatial Interferometer} (ISI, {\em Tuthill et
al.},~2002) have succeeded in resolving the brightest young systems on
smaller scales for the first time, establishing MIR sizes of $\sim
10$s~AU.  Although the observed samples are still small, and thus far
concern mostly HAe objects, these initial measurements have already
provided a number of interesting, albeit puzzling results: MIR
characteristic sizes may not always be well predicted by simple
(flared) disk models (some objects are ``under-sized'' compared to
these predictions, opposite to the NIR result), and may not always
connect in a straightforward way with measurements at shorter (NIR)
and longer (mm) wavelengths.

These pioneering efforts were however limited in resolution (for the
single-aperture techniques) or sensitivity (for the heterodyne ISI),
and probing the MIR emission on small scales has only recently become
possible with the advent of the new-generation long baseline MIR
instruments (VLTI/MIDI, {\em Leinert et al.},~2003; KI/Nuller, {\em Serabyn
et al.},~2004).

Two main factors determine the interferometric signature of a dusty
protoplanetary disk in the MIR: The overall geometry of the disk, and
the composition of the dust in the disk ``atmosphere.'' The most
powerful way to distentangle these two effects is via spectrally
resolved interferometry, and this is what the new generation of
instruments such as MIDI at the VLTI are capable of providing.

\begin{figure}[htbp]
\epsscale{1.0}
\plotone{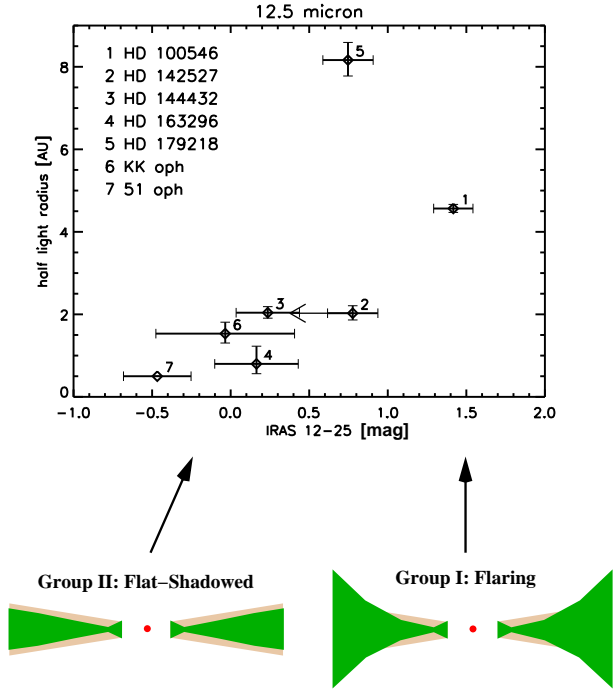}
\caption{\small 
Relation between $12.5 \, \mu m$ sizes (measured as half-light radius) and
infrared slope (measured by IRAS colors) for the first HAeBe
stars observed with MIDI on the VLTI.  This size-color relation is
consistent with expectations from the SED classification of {\em Meeus
et al.}~(2001) into objects with flared {\em vs.} flat outer disks. 
\label{mirsizes-fig}}
\end{figure}

First results establishing the MIR sizes of young circumstellar disks
were obtained by VLTI/MIDI for a sample of seven HAeBe objects ({\em
Leinert et al.}~2004). The characteristic sizes measured, based on
visibilities at the wavelength of $12.5 \, \mu m$, are in the range
1--10~AU. Moreover, as shown in Fig.~\ref{mirsizes-fig}, they are
found to correlate with the IRAS~[12]--[25] color, with redder objects
having larger MIR sizes.  These observations lend additional support
to the SED-based classification of {\em Meeus et al.}~(2001) and {\em
Dullemond and Dominik}~(2004), whereby redder sources correspond to
disks with a larger degree of flaring, which therefore also emit
thermally in the MIR from larger radii, compared to sources with
flatter disks.

A more powerful analysis is possible by exploiting the spectral
resolution capabilities of the MIDI instrument. Based on the reddest
object in the {\em Leinert et al.}~2004 sample (HD~100546), the
spectrally resolved visibilities also support the {\em Meeus et
al.}~(2001) classification: this object displays a markedly steeper
visibility drop between $8-9 \, \mu m$, and lower visibilities at
longer wavelengths than the rest of the objects in the sample,
implying the expected larger radius for the MIR emission. These first
observations have also been used to test detailed disk models that
were originally synthesized to fit the SEDs of individual objects
({\em Dominik et al.},~2003); without any feedback to the models {\em
Leinert et al.}~(2004) find encouraging qualitative agreement between
the predicted spectral visibility shapes and the interferometer data.

It remains to be seen whether direct fitting to the interferometer
data will solve the remaining discrepancies (see e.g., the work of
{\em Gil et al.},~2005 on the somewhat unusual object 51~Oph) .  In
general, changes in both disk geometry and dust composition strongly
affect the MIR spectral visibilities. However simulations by {\em van
Boekel et al.}~(2005) have shown that the problem is tractable and
even just a few well-chosen baselines (and wavelengths) permit a test
of some of the key features of recently proposed disk models, such as
the presence of the putative puffed-up inner rim and the degree of
outer disk flaring (which in these models is influenced by shadowing
by the inner rim).
%Thus far, a parametrised disk model satisfying both the
%spectral and interferometric constraints has been presented only for
%the somewhat unusual object 51~Oph ({\em Gil et al.},~2005), but the
%solution found requires a high accretion rate inconsistent with other
%evidence, signaling the need for further physical characterization.

%From the slope of the sub-mm/mm SED (and under the assumption that the
%disks are optically thin at these wavelengths), it has been shown that
%HAe and late HBe objects with smaller MIR flux excess have larger
%mid-plane dust grains, perhaps implying an evolutionary sequence of
%flared to self-shadowed disks due to grain growth and settling; or may
%alternatively indicate enhanced small-particle replenishment for the
%flared systems ({\em Acke et al.},~2004).

Clearly, the spectrally and spatially resolved disk observations made
possible by instruments such as VLTI/MIDI will prove crucial towards
further exploring exciting prospects for linking the overall disk
structure, the properties of mid-plane and surface layer dust, and
their time evolution towards the planet-building phase (see also the
chapter by {\em Dominik et al.}).

%%%%%%%%%%%%%%%%%%%%%%%%%%%%%%%%%%%%%%%%%%%%%%%%%%%%%%%%%%%%%%%%%%%%%%%

%%%%%%%%%%%%%%%%%%%%%%%%%%%%%%%%%%%%%%%%%%%%%%%%%%%%%%%%%%%%%%%%%%%%%%%

%%%%%%%%%%%%%%%%%%%%%%%%%%%%%%%%%%%%%%%%%%%%%%%%%%%%%%%%%%%%%%%%%%%%%%%

\bigskip
\noindent
\textbf{ 3.2  Dust Mineralogy and Mixing}
\bigskip

% do 10um RYTau results fit the PTI results?

The MIR spectral region contains strong resonances of abundant dust
species, both O--rich (amorphous or crystalline silicates) and C--rich
(Polycyclic Aromatic Hydrocarbons, or PAHs). Therefore, MIR
spectroscopy of optically thick proto-planetary disks offers a rich
diagnostic of the chemical composition and grain size of dust in the
upper disk layers or ``disk atmosphere''.  At higher spectral
resolution, $R \sim$~few~100, gas emission such as [NeII] lines and
lines of the Hund and higher hydrogen series can be observed.  It
should be noted that these observations do not constrain the
properties of large ($> 2-4 \, \mu m$) grains, since these have little
or no spectral structure at MIR wavelengths. In the sub-micron and
micron size range however, different silicate particles can be well
distinguished on the basis of the shape of their emission features.
Clearly, valuable information can be obtained from spatially
unresolved observations -- the overall SED constrains the distribution
of material with temperature (or radius), and MIR spectroscopy
provides the properties of dust. However, strong radial gradients in
the nature of the dust are expected, both in terms of size and
chemistry, that can only be observed with spatially resolved
observations.
%With spatially resolved observations, this type of dust processing can
%be searched at different spatial scales in the disk, rather than for
%the (unresolved) system as a whole.

{\em Van Boekel et al.}~(2004) have demonstrated the power of
spectrally resolved MIR interferometry, by spatially resolving three
proto-planetary disks surrounding HAeBe stars across the N-band.  The
correlated spectra measured by VLTI/MIDI correspond to disk regions at
1--2~AU. By combining these with unresolved spectra, the spectrum
corresponding to outer disk regions at 2--20~AU can also be deduced.
These observations have revealed radial gradients in dust
crystallinity, particle size (grain growth), and at least in one case
(HD~142527) chemical composition. These early results have revived the
discussion of radial and out-of-the-plane mixing.

Interstellar particles observed in the direction of the galactic
center are mainly small amorphous Olivine grains
(Fe$_{2-2x}$Mg$_{2x}$SiO$_4$) with some admixture of Pyroxene grains
(Fe$_{1-x}$Mg$_{x}$SiO$_3$) ({\em Kemper et al.},~2004).  In our
planetary system, comets like Halley, Levi or Hale-Bopp ({\em Hanner
et al.},~1994; {\em Crovisier et al.},~1997) show the features of
crystalline Forsterite (an Mg-rich Olivine, x=1), in particular the
conspicuous emission at $11.3 \, \mu m$.

\begin{figure}[!h]
\epsscale{1.0}
\plotone{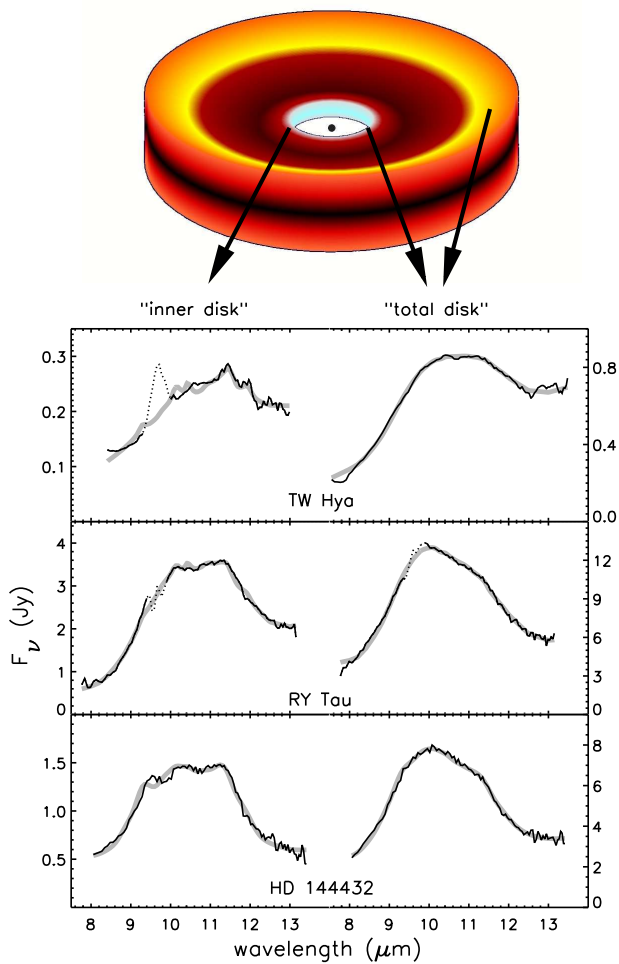}
%\rule{5pt}{2cm}
\caption{\small 
Evidence for radial changes of dust composition in the circumstellar
disks of young stars. Infrared spectra of the inner $\approx$~1--2~AU
are shown in the left panels, and of the total disk
($\approx$~1--20~AU) in the right panels. The different shapes of the
silicate emission lines testify that in the inner disk -- as probed by
the interferometric measurement -- the fraction of crystalline
particles is high while the fraction of small amorphous particles is
significantly reduced.  TW~Hya (0.3~L$_{\sun}$) and RY~Tau
(18~L$_{\sun}$) approximately bracket the luminosity range of T~Tauri
stars, the 10~L$_{\sun}$ HAe star HD~144432 (from {\em van Boekel et
al.},~2004) is included to emphasize the similarity between these two
classes of young objects. Credit: Roy van Boekel.\label{mirgradients-fig}}
\end{figure}

The VLTI/MIDI observations have revealed that crystallinization in
young circumstellar disks can be a very efficient process, by
suggesting a very high fraction of crystalline dust in the central
1--2~AU; and that the outer 2--20~AU disk posseses a markedly lower
crystalline fraction, but still much higher than in the interstellar
medium. Combined with the very young evolutionary status of these
systems (as young as 1~Myr), these observations imply that
proto-planetary disks are highly crystalline well before the onset of
planet formation.  The presence of crystalline dust in relatively cold
disk regions (or in Solar System comets) is surprising, given that
temperatures in these regions are well below the glass temperature of
$\sim 1000$~K, but can be explained by chemical models of
proto-planetary disks that include the effects of radial mixing and
local processes in the outer disk ({\em Gail},~2004; {\em
Bockel\'{e}e-Morvan et al.},~2002).  The radial gradient found in dust
chemistry is also in qualitative agreement with the predictions of
these models.  Finally, the width of the silicate feature in the MIDI
spectra corresponding to the inner and outer disk have revealed a
radial gradient in (amorphous) grain size, with small grains (broader
feature) being less abundant in the inner disk regions.  This is
perhaps expected, given that grain aggregation is a strong function of
density (see the chapter by {\em Natta et al.}).
Fig.~\ref{mirgradients-fig} includes an example illustrating radial
dust mineralogy gradients for the HAe object HD~144432.

These radial gradients in dust mineralogy are of course not restricted
to HAeBe stars, but are expected in all YSO disks.  In
Fig.~\ref{mirgradients-fig} we show preliminary VLTI/MIDI results for
two low-mass objects of different evolutionary status, TW~Hya and
RY~Tau ({\em Leinert and Schegerer},~2006, in preparation). For 
TW~Hya, the spatially unresolved N-band spectrum (right panel) shows
no strong evidence for crystalline silicates, while the correlated
flux spectrum (left panel) shows a weak but highly processed silicate
band. Preliminary modelling indicates that the inner disk region of
TW~Hya contains a high fraction of crystalline silicates (30\%), and
that the amorphous grains have aggregated to larger units. As with the
HAeBe stars, not all T~Tauri objects show the same features; e.g., the
effect is qualitatively similar but quantitatively less pronounced in
RY~Tau, as can be seen in Fig.~\ref{mirgradients-fig}.

%From unresolved spectrophotometric observations (e.g. {\em Przygodda
%et al.},~2003) it appears that in general the crystallinity of T~Tauri
%disks is somewhat lower than that of HAeBe stars, although the reasons
%for this difference are unclear at present. As for the HAeBe objects,
%the unresolved N--band flux spectra of T~Tauri stars have very
%different shapes, often indicating the removal of small amorphous
%silicate grains from the inner disk region. However, spectral features
%of small grains in the disk surface are also typical of some of these
%systems. Given the stellar ages these small grains should have settled
%to the disk mid-plane and disappeared.  Their presence perhaps
%indicates strong turbulence-driven vertical mixing. Although unclear
%at present, the precise mechanisms may be further elucidated with
%detailed targeted interferometric observations of selected objects.

%%%%%%%%%%%%%%%%%%%%%%%%%%%%%%%%%%%%%%%%%%%%%%%%%%%%%%%%%%%%%%%%%%%%%%%

%%%%%%%%%%%%%%%%%%%%%%%%%%%%%%%%%%%%%%%%%%%%%%%%%%%%%%%%%%%%%%%%%%%%%%%

%\input{ysos_table}

%%%%%%%%%%%%%%%%%%%%%%%%%%%%%%%%%%%%%%%%%%%%%%%%%%%%%%%%%%%%%%%%%%%%%%%

\bigskip
\centerline{\textbf{ 4. OTHER PHENOMENA}}
\bigskip

\bigskip
\noindent
\textbf{4.1 Outflows and Winds}
\bigskip

The power of spectrally resolved interferometric measurements has
recently become available in the NIR spectral range as well
(VLTI/AMBER, {\em Malbet et al.},~2004; CHARA/MIRC, {\em Monnier et
al.},~2004a).  In addition to providing the detailed wavelength
dependence of inner disk continuum (dust) emission, these capabilities
enable detailed studies of the physical conditions and kinematics of
the gaseous components in which emission and absorption lines arise
(e.g., Br$_\gamma$, CO and H$_2$ lines; as probes of hot winds, disk
rotation and outflows, respectively).

As a rather spectacular example of this potential, during its first
commissioning observations the VLTI/AMBER instrument spatially
resolved the luminous HBe object MWC~297, providing visibility
amplitudes as a function of wavelength at intermediate spectral
resolution $R=1500$ across a $2.0-2.2 \, \mu m$ band, and in
particular a finely sampled Br$_\gamma$ emission line ({\em Malbet et
al.},~2006). The interferometer visibilities in the Br$_\gamma$ line
are $\sim 30$\% lower than those of the nearby continuum, showing that
the Br$_\gamma$ emitting region is significantly larger than the NIR
continuum region.

Known to be an outflow source ({\em Drew et al.},~1997), a preliminary
model has been constructed by {\em Malbet et al.}~(2006) in which a
gas envelope, responsible for the Br$_\gamma$ emission, surrounds an
optically thick circumstellar disk (the characteristic size of the
line emitting region being 40\% larger than that of the NIR disk).
This model is succesful at reproducing the new VLTI/AMBER measurements
as well as previous continuum interferometric measurements at shorter
and longer baselines ({\em Millan-Gabet et al.},~2001; {\em Eisner et
al.},~2004), the SED, and the shapes of the H$_\alpha$, H$_\beta$ and
Br$_\gamma$ emission lines.

The precise nature of the MWC~297 wind however remains unclear; the
limited amount of data obtained in these first observations can not,
for example, discriminate between a stellar or disk origin for the
wind, or between competing models of disk winds (e.g., {\em Casse and
Ferreira},~2000; {\em Shu et al.},~1994). These key questions may
however be addressed in follow-up observations of this and similar
objects, perhaps exploiting enhanced spectral resolution modes (up to
$R=10000$ possible with VLTI/AMBER) and closure phase capabilities.

{\em Quirrenbach et al.}~(2006) have also presented preliminary
VLTI/MIDI results of rich spectral content for the wind environment in
another high luminosity YSO, MWC~349--A. The general shape of the
N-band spectrally resolved visibilities is consistent with disk
emission, as in the {\em Leinert et al.}~(2004) results
(Section~3.1). In addition, the visibility spectrum contains
unprecedented richness, displaying over a dozen lines corresponding to
wind emission in forbidden ([NeII], [ArIII], [SIV]) and H
recombination lines. The relative amplitudes of the visibilities
measured in the continuum and in the lines indicate that the forbidden
line region is larger than the dust disk, and that the recombination
region is smaller than the dust disk (at least along some directions).
Moreover, differential phases measured with respect to the continuum
also display structure as a function of wavelength, showing clear
phase shifts that encode information about the relative locations of
the emitting regions for the strongest emission lines.

%Substantial modelling efforts is required to fully exploit the
%multitude of observables provided by, ideally, the combination of
%these new-generation instruments. It is to be expected that the
%resulting new insight will play a key role in elucidating the
%interplay of outflow and disk phenomena in these luminous young stars.

\bigskip
\noindent
\textbf{ 4.2 Multiplicity and Stellar Masses}
\bigskip

Most young (and main sequence) stars are members of multiple systems,
likely as a result of the star formation process itself (see the
chapters by {\em Goodwin et al.} and by {\em Duchene et al.}).  The
measurement of the physical orbits of stars in multiple systems
provide the only direct method for measuring stellar masses, a
fundamental stellar parameter. In turn, by placing stars with measured
masses in an HR-diagram, models of stellar structure and evolution can
be critically tested and refined, provided the mass measurements have
sufficient accuracy (see the chapter by {\em Mathieu et al.}). Generally
speaking, dynamical and predicted masses agree well for $1-10
M_{\sun}$ stars in the main sequence. However, fundamental stellar
properties are much less well known for pre-main sequence (PMS) stars,
particularly of low-mass ({\em Hillenbrand and White},~2004), and call
for mass measurements with better than 10\% accuracy.

Optical interferometers can spatially resolve close binaries (having
separations of order $\sim 1$~mas, or 1/10ths~AU at 150~pc), and
establish the apparent astrometric orbits, most notably its
inclination (for eclipsing binaries, the inclination is naturally well
constrained). In combination with radial velocity measurements using
Doppler spectroscopy, a full solution for the physical orbit and the
properties of the system can then be obtained, in particular the
individual stellar masses and luminosities.  This method has proved
very fruitful for critical tests of stellar models for non-PMS stars
(e.g., {\em Boden et al.},~2005a), however few simultaneous
radial velocity and astrometric measurements exist for PMS stars.

{\em Boden et al.}~(2005b) performed the first direct measurement of PMS
stellar masses using optical interferometry, for the the double-lined
system HD~98800--B (a pair in a quadruple system located in the
TW~Hya association). Using observations made at the KI and by the Fine
Guidance Sensors aboard the Hubble Space Telescope, and in combination
with radial velocity measurements, these authors establish a
preliminary orbit which allowed determination of the (sub-solar)
masses of the individual components with 8\% accuracy. Comparison with
stellar models indicate the need for sub-solar abundances for both
components, although stringent tests of competing models will only
become possible when more observations improve the orbital phase
coverage and thus the accuracy of the stellar masses derived.

%Other PMS systems being pursued for which accurate masses may be
%obtained include: MWC~361--A at the high mass end of the spectrum
% (likely two Be stars, {\em Millan-Gabet et al.},~2001; {\em Monnier et
%al.},~2006, in preparation); Haro~1--14C ({\em Schaefer et al.},~2005;
%and see also the chapter by {\em Mathieu et al.}); HD~98800--A ({\em Boden
%et al.},~2005); and the triple T~Tauri system GW~Ori (also believed to
%contain circumstellar and circumbinary disks, and for which {\em
%Berger et al.}~(2006) present preliminary separation vectors and the
%first reconstructed image of a young multiple system).

Naturally, disk phenomena (circumstellar and/or circumbinary) are also
often observed in young multiple systems (see the chapter by {\em Monin et
al.}); and a number of recent observations also address this
interesting and more complicated situation.  Indeed, successful
modelling of the individual HD~98800--B components by {\em Boden et
al.}~(2005b) requires a small amount of extinction (A$_V$ = 0.3) toward
the B--components, not required toward the A--components, and possibly
due to obscuration by circumbinary disk material around the B system
(originally hypothesized by {\em Tokovinin},~1999). As another
example, based on a low-level oscillation in the visibility amplitude
signature in the PTI data for FU~Ori, {\em Malbet et al.}~(2005) claim
the detection of an off-centered spot embedded in the disk that could
be physically interpreted as a young stellar or proto-planetary
companion (located at $\sim 10$~AU), and possibly be at the origin of
the FU~Ori outburst itself.

%In summary, the improved infrared sensitivity of new interferometer
%facilities has enabled the first measurements of PMS binary orbits and
%stellar masses, and promises to result in establishing fundamental
%stellar properties for a larger sample of the closest systems across
%the mass range, and in critical tests of state of the art stellar
%models. 
As noted by {\em Mathieu}~(1994), spectroscopic detection of radial
velocity changes is challenging in the presence of strong emission
lines, extreme veiling or rapid rotation, all of which are common in
PMS stars. Therefore, imaging techniques are unique discovery tools
for PMS multiples. For the closest systems in particular,
interferometric techniques can reveal companions with separations
$\sim \times 10$ smaller than single-aperture techniques (e.g.,
speckle or adaptive optics). Therefore, the increasing number of new
detections will also add an important new sample to
statistical studies aimed at determining the multiplicty fraction for
young stars, its dependence on separation, stellar properties and
evolutionary status, and implications for circumstellar disk survival
in close binary environments.

\bigskip
\centerline{\textbf{ 5. SUMMARY AND OPEN QUESTIONS}}
\bigskip

\bigskip
\noindent
\textbf{ 5.1. Summary}
\bigskip

Long-baseline interferometers operating at near and mid-infrared
wavelengths have spatially resolved the circumstellar emission of a
large number of YSOs of various types (60 objects published to date),
most within the last few years.  While the new observables are
relatively basic in most cases (broadband visibility amplitudes), they
have provided fundamentally new direct information (characteristic
sizes and crude geometry), placing powerful new constraints on the
nature of these circumstellar environments. In addition, more powerful
observational capabilities combining high spatial resolution and
spectral resolution have just begun to deliver their potential as
unique tools to study YSO environments in exquisite detail and to
provide breakthrough new molecular and kinematic data.

The principal well-established observational facts may be summarized
as follows:

{\em (i)}
For HAe, late HBe and T~Tauri objects the measured NIR
characteristic sizes are much larger (by factors of $\sim 3 - 7$) than
predicted by previous-generation models which had been reasonably
successful at reproducing the unresolved spectro-photometry. In
particular, the measured sizes are larger than initial inner dust hole
radii estimates and than magnetospheric truncation radii. The measured
NIR sizes strongly correlate with the central luminosity, and the
empirical NIR size {\em vs.} luminosity relation
(Fig.~\ref{size-fig}) suggests that the NIR emission is located at
radii corresponding to sublimation radii for dust directly heated by
the central star.  These measurements have motivated in part the
development of a new class of models for the origin of the NIR disk
emission (the ``puffed-up'' inner wall models) which not only provide
qualitative agreement with the interferometer data, but also solves
the SED ``NIR bump'' problem for HAe objects.

{\em (ii)} Some (but not all) of the earliest-type HBe objects are
different than the later types in terms of their NIR size-scales. Many
are inferred to have relatively small inner dust holes in better
agreement with the classical disk picture, while others have larger
NIR sizes consistent with the inner rim models (or perhaps even
larger than predictions of either competing disk model,
e.g., LkH$\alpha$~101).

{\em (iii)} FU~Ori, the prototype for the sub-class of YSOs expected
to have disks which dominate the total luminosity, is well described
by the canonical model temperature law (based on modelling the NIR
interferometry data and SED). For other systems in this class however,
the observations reveal more complexity and the need for data at
multiple baselines and wavelengths in order to discriminate between
competing scenarios.

{\em (iv)} Young disks have also been resolved at MIR wavelengths,
revealing characteristic sizes which correlate with red color (IRAS
$12-25 \, \mu m$). This link appears to support an SED-based
classification based on the degree of outer disk flaring, although the
current sample of measured MIR sizes is too small to extract a firm
conclusion.

{\em (v)} Spectrally-resolved MIR visibilities have emerged as a
powerful tool to probe the dust minerology in young disks.  Silicate
dust appears to be deficient in small ($0.1 \, \mu m$) amorphous
grains and more crystalline (less amorphous) close to the central star
when compared to the disk as a whole.

{\em (vi)} The power of spectrally-resolved visibilities, both NIR and MIR,
has also been dramatically demonstrated in observations of lines
corresponding to the gas component in wind phenomena associated with
high-mass young stars.

{\em (vii)} Masses have been measured for young stars for the first
time using interferometric techniques, in combination with
spectroscopic radial velocity techniques; and a number of systems are
being pursued with the primary goal of providing useful constraints to
state of the art models of stellar structure and evolution.

\bigskip
\noindent
\textbf{ 5.2. Open Questions}
\bigskip

The new interferometric observations have allowed a direct view of the
inner regions of YSO accretion disks for the first time.  While this
work has motivated significant additions to our standard picture of
these regions (such as the puffed-up inner rim and spatially-varying
dust properties), continued theoretical development will require more
sophisticated and complete interferometry data.  In this section, we
outline the most pressing un-resolved issues exposed by the progress
describe above:

{\em (i)} A disk origin for the resolved NIR emission has not been
unambiguously established.  Most notably, {\em Vincovik et al.}~(2006)
(and references therein) consider that the remaining scatter in the
size-luminosity relation for HAeBe objects indicates that the
bright-ring model does not adequately describe all the objects with a
uniform set of parameters, and propose instead a disk plus compact
halo model, where the resolved NIR emission in fact corresponds to the
compact halo component.  Because the halo need not be spherically
symmetric, the elongations detected at the PTI do not rule it
out. Effectively re-igniting the {\em disk vs. envelope} debate of the
previous decade (see e.g., {\em Natta et al.},~2000), {\em Vincovik et
al.}~(2006) also show that this model equivalently reproduces the SED,
including the NIR bump.

{\em (ii)} The detailed properties of the putative puffed-up inner rim
remain to be established. Assumptions about the grain size and density
directly impact the expected rim location (dust sublimation radius),
and based on size comparisons (Fig.~\ref{size-fig}) current
observations appear to indicate that a range of properties may be
needed to explain all disks, particularly among the T~Tauri objects.
HAe objects have NIR sizes consistent with optically thin dust and
relatively large grains ($\sim 1 \, \mu m$).  The lowest luminosity
T~Tauri objects however favor either optically thick dust or smaller
grains, or a combination of both. Currently however, the T~Tauri
sample is still small, several objects are barely resolved, and part
of the scatter could be due to unreliable SED decomposition due to
photometric variability and uncertain stellar properties. Detailed
modelling of individual objects and crucial supporting observations
such as infrared veiling (to directly establish the excess flux) are
needed to help resolve these uncertainties.

{\em (iii)} Using IR spectroscopy, it has been established that the
{\em gas} in young disk systems extends inward of the
interferometrically deduced inner dust radii (see the chapter by {\em
Najita et al.}).  Whether or not magnetospheric accretion theories can
accomodate these observations remains to be explored.  We note that
current interpretations of the CO spectroscopy do not take into
account the inner rim, relying instead on previous-generation models
that have been all but ruled out for the inner few~AU. Finally,
effects of NIR emission by gas in the inner disk, both on the measured
visibilities and on the SED decomposition (both of which affect the
inferred NIR sizes), need to be quantified in detail on a case-by-case
basis.

% {\em (v)} The effects of NIR emission by gas in the inner disk and of NIR
%scattering by large scale dust, both on the measured visibilities and
%on the SED decomposition (both of which affect the inferred NIR
%sizes), need to be quantified in detail on a case-by-case basis.

{\em (vi)} Large scale ($\sim 50 - 1000$~mas) ``halos'' contributing small
but non-negligible NIR flux ($\sim 10$\%) are often invoked as an
additional component needed to satisfactorily model some YSO disks
(e.g., the FU~Orionis objects, or most recently in the IOTA-3T HAeBe
sample of {\em Monnier et al.},~2006, in preparation). The origin of
this halo material is however unclear, as is its possible connection
with the compact halos proposed by {\em Vincovik et al.}~(2006).  The
observational evidence should motivate further work on such
multi-component models.

{\em (vii)} Late (HAe, late HBe) and early HBe systems appear to have
different NIR scale properties (see Section~2.1), although there are
notable exceptions, and the sample is small. Differences between these
types have also been noted in terms of their mm emission (absent in
most early HBe, in contrast to the late Herbig and T~Tauri objects),
perhaps due to outer-disk photoevaporation ({\em Hollenbach et
al.},~2000, and see also the chapter by {\em Dullemond et
al.}). Whether the observed differences in inner disk structure are
due to different accretion mechanisms (disk accretion
vs. magnetospheric accretion) or to gas opacity effects remains to be
investigated.

{\em (viii)} Do all FU~Orionis objects conform to the canonical
accretion disk model? If so, the additional question arises as to why
their temperature structure would be that simple. Their light curves
are fading, indicating a departure from a steady-state disk. Indeed,
the high outburst accretion rates are expected to decay, and to do so
in a radially dependent manner, leading to non-standard temperature
laws.  First observations using VLTI/MIDI do not indicate a simple
connection between the NIR and MIR radial disk structure. Indeed,
fitting the MIR visibilities and SED of V1647~Ori requires a very flat
$q=-0.5$ radial temperature exponent ({\em \'{A}brah\'{a}m et
al.},~2006), and preliminary analysis of FU~Ori itself ({\em
Quanz},~2006) indicates that the model constructed by {\em Malbet et
al.}~(2005) to fit the NIR AMBER data and the SED does not reproduce
(over-estimates) the measured N-band visibilities.

{\em (ix)} The overall disk structure needs to be secured.  While the NIR
data probe primarily the hottest dust, the MIR observations are
sensitive to a wider range of temperatures.  Given the complexity of
the disk structure under consideration, correctly interpreting the MIR
and NIR data for the same object is proving challenging. Assumptions
made for modelling the NIR data, the MIR dust features, and CO line
profiles are generally not consistent with each other.  Excitingly,
the joint modeling of near and mid-IR high resolution data can yield
the temperature profile of the disk surface layers.

{\em (x)} Chemical models of proto-planetary disks, including the effects
of radial transport and vertical mixing, must explain the observed
dust mineralogy, and the differences seen among different types of
objects. Also, we note that the effect of the inner rim (which
contributes $\sim 20$\% of the MIR flux -- see
Fig.~\ref{radial-fig}) on the interpretation of the observed
Silicate features and inferred mineralogy gradients must be assesed in
detail.

\bigskip
\centerline{\textbf{ 6. FUTURE PROSPECTS}}
\bigskip

This review marks the maturation of two-telescope (single-baseline)
observations using infrared interferometry.  However, much more
information is needed to seriously constrain the next generation of
models (and the proliferating parameters) and to provide reliable
density and temperature profiles as initial conditions for planet
formation theories. Fortunately, progress with modern interferometers
continues to accelerate and will provide the unique new observations
that are needed.  In this section, we briefly discuss the expected
scientific impact on yet-unexplored areas that can be expected from
longer baselines, multi-wavelength high-resolution data, spectral line
and polarization capabilities, closure phase data and aperture
synthesis imaging.  We will end with suggestions for disk
modelers who wish to prepare for the new kinds of interferometric data
in the pipeline.

\bigskip
\noindent
\textbf{ 6.1. Detailed Disk Structure}
\bigskip

All interferometric observations of YSO disks to date have used
baselines $\lesssim 100$~meters, corresponding to angular resolution
of $\sim$~2~and~11~milliarcsec at 2.2~and~10~microns respectively.
While sufficient to resolve the overall extent of the disks, longer
baselines are needed to probe the internal structure of this hot dust
emission. For instance, current data can not simultaneously constrain
the inner radius and thickness of the inner rim dust emission, the
latter often assumed to be ``thin'' ($10-25$\% of radius).  Nor can
current data independently determine the fraction of light from the
disk compared to star (critical input to the models), relying instead
on SED decomposition.

Longer baselines from the CHARA Interferometer (330~meters) and from
VLTI Interferometer (200~meters) can significantly increase the
angular resolution and allow the rim emission to be probed in detail.
For instance, if the NIR emission is contained in a thin ring, as
expected for ``hot inner wall'' models, there will be a dramatic
signal in the second lobe of the interferometer visibility response.
Competing ``halo'' models predict smoother fall-offs in brightness
(and visibility) and thus new long-baseline data will provide further
definitive and completely unique constraints on the inner disk
structure of YSOs (first long-baseline CHARA observations of YSOs were
presented by {\em Monnier et al.},~2005b).

While even two-telescope visibility data provide critical constraints,
ultimately we strive for aperture synthesis imaging, as routinely done
at radio wavelengths.  The first steps have recently been taken with
closure phase results on HAeBe objects using IOTA/IONIC3 ({\em
Millan-Gabet et al.},~2006b; {\em Monnier et al.},~2006; both in
preparation).  Closure phases are produced by combining light from
three or more telescopes, and allow ``phase'' information to be
measured despite atmospheric turbulence (see e.g., {\em
Monnier},~2000). While this phase information is essential for image
reconstruction, the closure phase gives unambiguous signal of
``skewed'' emission, such as would arise from a flared disk viewed
away from the pole.  ``Skew'' in this context indicates a deviation
from a centro-symmetric brightness.

Closure phases are interesting for measuring the vertical structure of
disks.  Early hot inner wall models (e.g., {\em Dullemond et
al.},~2001) posited vertical inner walls.  Recently, the view was made
more realistic by {\em Isella and Natta}~(2005) by incorporating
pressure-dependent dust sublimation temperatures to curve the inner
rim away from the midplane.  Closure phases can easily distinguish
between these scenarios because vertical walls impose strong skewed
emission when a disk is viewed at intermediate inclination angles,
while curved inner walls appear more symmetric on the sky unless
viewed nearly edge-on.  NIR closure phase data will soon be available
from IOTA/IONIC3, VLTI/AMBER and CHARA/MIRC that can be used to
measure the curvature and inner rim height through model fitting of
specific sources.

Aperture synthesis imaging of selected YSOs using the VLTI and CHARA
arrays should also be possible within the next five years.  By
collecting dozens of closure phase triangles and hundreds of
visibilities, simple images can be created as has been demonstrated
for LkH$\alpha$~101 using the aperture masking technique.  First steps
in this direction have been presented for the triple T~Tauri system
GW~Ori (also believed to contain circumstellar and circumbinary disks)
by {\em Berger et al.}~(2006), who present preliminary separation
vectors and the first reconstructed image of a young multiple
system. Ultimately, YSO imaging will allow the first model-independent
tests of disk theories; all current interpretations are wed to
specific (albeit general) models and the first images may be
eye-opening.

As described earlier (Section~3), it is essential to know the
properties of circumstellar dust particles in YSO disks in order to
interpret a broad range of observations, most especially in standard
SED modelling.  Interferometers can measure dust properties, such as
size distribution and composition, using the combination of NIR, MIR,
and polarization measurements.  Efforts are being explored at the PTI,
IOTA, VLTI and CHARA interferometers to measure the disk emission in
linearly polarized light (see {\em Ireland et al.},~2005 for similar
first results for dust around AGB stars).  Because scattering is
heavily dependent on grain-size, the polarization-dependent size
estimates at different wavelengths will pinpoint the sizes of the
grains present in the inner disk.

With its nulling capability at MIR wavelengths, the KI ({\em Serabyn
et al.},~2004) will soon address a wide range of YSO phenomena, from
spectrally resolved MIR disk structure in young disks to exo-zodiacal
emission ({\em Serabyn et al.},~2000), the latter being a crucial
element in the selection of favorable targets for future space-based
planet finding missions (e.g., Terrestrial Planet Finder/Darwin, {\em
Fridlund},~2003).

Proto-planets forming in circumstellar disks are expected to carve
fine structure, such as gaps, opening the possibility of directly
detecting this process via high dynamic range interferometric
techniques (e.g., highly accurate visibility amplitude
measurements). While some initial simulations are pessimistic that new
interferometers can resolve such disk structures ({\em Wolf et
al.},~2002), it is clear that theorists have only begun to investigate
the many possibilities and interferometers are gearing up to meet the
necessary observational challenges ahead.

Finally, although outside the scope of this review, we point out the
tremendous potential of the upcoming Atacama Large Millimiter Array
(ALMA, {\em Wootten},~2003). In particular, the combination of
infrared, mm and sub-mm interferometry will probe the entire YSO disk
at all scales.

\bigskip
\noindent
\textbf{ 6.2. Dynamics}
\bigskip

New spectral line (e.g., VLTI/AMBER) capabilities have exciting
applications.  For instance, spectro-interferometry can measure the
Keplerian rotation curves (in CO) for disks to derive dynamical masses
of young stars as has been done using mm-wave interferometry ({\em
Simon et al.},~2000). Clearly, the potentially very powerful
combination of spectroscopy and interferometry has only begun to be
explored.

The dynamic time scale for inner disk material can be $\lesssim
1$~year, thus we might expect to see changes with time. The higher the
angular resolution, the greater the possibility that we can identify
temporal changes in the circumstellar structures.  If disk
inhomogeneities are present, new measurements will discover them and
track their orbital motions.  Evidence of inner disk dynamics has been
previously inferred from reflection nebulosity (e.g., HH~30, {\em
Stapelfeldt et al.},~1999) and photometric variability (e.g., {\em
Hamilton et al.},~2001); but only recently directly imaged at
AU~scales (LkH$\alpha$~101, {\em Tuthill et al.},~2002).

\bigskip
\noindent
\textbf{ 6.3. The Star - Disk - Outflow Connection}
\bigskip

The interface between the star, or its magnetosphere, and the
circumstellar disk holds the observational key into how disks mediate
angular momentum transfer as stars accrete material.  The relevant
spatial scales (1 to few~$R_\star$) may be resolvable by the longest
baselines available, providing unique tests of magnetospheric
accretion theories (see the chapter by {\em Bouvier et al.}).

In addition to being surrounded by pre-planetary disks, YSOs are often
associated with outflow phenomena, including optical jets, almost
certainly associated with the disk accretion process itself (see the
chapter by {\em Bally et al.}).  The precise origin of the jet
phenomena is however unknown, as is the physical relation between the
young star, the disk, and jets (see the chapter by {\em Ray et al.}).
Discriminating among competing models, e.g., jets driven by external
large scale magnetic fields (the ``disk wind'' models or stellar
fields models) or by transient local disk fields, can be helped by
determining whether the jets launch near the star, or near the disk.
Observationally, progress depends in part on attaining sufficient
spatial resolution to probe the jet launching region directly, a
natural task for optical interferometers (e.g., {\em Thiebaut et
al.},~2003).

\bigskip
\noindent
\textbf{ 6.4. New Objects}
\bigskip

%Essentially unexplored areas concern types of objects for which little
%or no observations have been carried out to date. We point out here
%important astrophysical questions concerning types of objects for
%which the observational effort has just begun, or can be expected to
%flourish in the next few years.

Infrared interferometry can directly address important questions
concerning the early evolution of high mass (O--B) protostars (see the
chapter by {\em Cesaroni et al.} for a review of the pressing issues), and
observational work in this area using infrared interferometry has just
begun (with the VLTI/MIDI observations of {\em Quirrenbach et
al.},~2006; {\em Feldt et al.},~2006). Most notably, it is not
presently known what role, if any, accretion via circumstellar disks
plays in the formation of high-mass stars. If present, interferometers
can resolve circumstellar emission, and determine whether or not it
corresponds to a circumstellar disk. Further, by studying disk
properties as a function of stellar spectral type, formation
timescales and disk lifetimes can be constrained.

Current studies have mainly targeted young disk systems, i.e. class~II
({\em Adams et al.},~1987).  More evolved, class~III, disks have
remained thus far unexplored (the only exception being V380~Ori in
{\em Akeson et al.},~2005a), due to the sensitivity limitations that
affected the first observations.  Moreover, directly establishing the
detailed disk structure of transition objects in the planet-building
phase and of debris disks on sub-AU to 10s~AU scales will likely be an
energetic area of investigation as the field progresses towards higher
dynamic range and spatial frequency coverage capabilities. First steps
have been taken in this direction, with NIR observations that resolve
the transition object TW~Hya ({\em Eisner et al.},~2006) and detect
dust in the debris disk around Vega ({\em Ciardi et al.},~2001; {\em
Absil et al.},~2006, in preparation).

% as was done for inner rim, size + sed can provide dust properties
% for these important transition objects.

Finally, we note that prospects also exist for the direct detection of
exo-planets from the ground, including interferometric techniques, and
we refer the reader to the chapter by {\em Beuzit et al.}

\bigskip
\noindent
\textbf{ 6.5. Comments on Modelling}
\bigskip

A number of new physical ingredients must be introduced into current
disk models in order to take advantage of interferometer data.  Disk
models should take into account both vertical variations (settling)
and radial dust differences (processing and transport).  Also, the
structure of inner wall itself depends on the physics of dust
destruction, and may depend on gas cooling and other physical effects
not usually explicitly included in current codes.  Perhaps codes will
need to calculate separate temperatures for each grain size and type
to accurately determine the inner rim structure.  Lastly, we stress
that the many free parameters in these models can only be meaningfully
constrained through a data-driven approach, fitting not only SEDs
(normal situation) but NIR and MIR interferometry simultaneously for
many individual sources.  With these new models, disk mass profiles
and midplane physical conditions will be known, providing crucial
information bridging the late stages of star formation to the initial
conditions of planet formation.

%%%%%%%%%%%%%%%%%%%%%%%%%%%%%%%%%%%%%%%%%%%%%%%%%%%%%%%%%%%%%%%%%%%%%%%

\bigskip

\textbf{ Acknowledgments.} 

We thank Roy van Boekel for providing Figs.~3~and~5 and for helpful
discussions.  We also thank many of our colleagues for illuminating
discussions while preparing this review, in particular Myriam Benisty,
Jean-Philippe Berger, Andy Boden, Kees Dullemond, Josh Eisner,
Lynne Hillenbrand and Andreas Quirrenbach.

\bigskip
%\newpage

\centerline\textbf{ REFERENCES}
\bigskip
\parskip=0pt
{\small
\baselineskip=11pt

\refs \'{A}brah\'{a}m P., K\'{o}sp\'{a}l \'{A}., Kun M., Mo\'{o}r A. et al.
(2004) \aap, {\em 428}, 89-97.

\refs \'{A}brah\'{a}m P., Mosoni, L., Henning Th., K\'{o}sp\'{a}l \'{A}.,
Leinert Ch.  et al. (2006) \aap, {\em in press}.

%\refs Acke B., van den Ancker M. E., Dullemond C. P., van Boekel R., 
%and Waters L. B. F. M. (2004) \aap, {\em 422}, 621-626.

\refs Adams F. C., Lada C. J., and Shu F. H. (1987) \apj, {\em 312}, 788-806.

\refs Akeson R. L., Ciardi D. R., van Belle  G. T., Creech-Eakman M. J., and Lada E. A.
(2000) \apj, {\em 543}, 313-317.

\refs Akeson R. L., Ciardi D. R., van Belle G. T., and Creech-Eakman, M. J.
(2002) \apj, {\em 566}, 1124-1131.

\refs Akeson R. L., Boden A. F., Monnier J. D., Millan-Gabet R., Beichman C.
et al. (2005a) \apj, {\em 635}, 1173-1181.

\refs Akeson R. L., Walker C. H., Wood K., Eisner J. A., and Scire E.
(2005b) \apj, {\em 622}, 440-450.

\refs Berger J.-P., Monnier J. D., Pedretti E., Millan-Gabet R., Malbet, F.
et al. (2005) in {\em Protostars and Planets V Poster Proceedings}
{\tt http://www.lpi.usra.edu/meetings/ \\ppv2005/pdf/8398.pdf}

\refs Bjorkman J. E. and Wood K. (2001) \apj, {\em 554}, 615-623.

\refs Boden A. F., Torres G., and Hummel C. A. (2005a) \apj, {\em 627}, 464-476

\refs Boden A. F., Torres G., and Hummel C. A. (2005b) \apj, {\em 627}, 464-476.

\refs Bockel\'{e}e--Morvan D., Gautier D., Hersant F., Hur\'{e}, J.-M., and Robert F.
(2002) \aap, {\em 384}, 1107-1118.

\refs Casse F. and Ferreira J. (2000) \aap, {\em 353}, 1115-1128.

\refs Chiang E. and Goldreich P. (1997) \apj, {\em 490}, 368.

\refs Ciardi D. R., van Belle G. T., Akeson R. L., Thompson, R. R., Lada, E. A.
et al. (2001), \apj, {\em 559}, 1147-1154.

\refs Colavita M. M., Wallace J. K., Hines B. E., Gursel Y., Malbet F.
et al. (1999) \apj, {\em 510}, 505-521.

\refs Colavita M. M., Akeson R., Wizinowich P., Shao M., Acton S. 
et al. (2003) \apj, {\em 529}, L83-L86.

\refs Colavita M. M., Wizinowich P. L., and Akeson R. L. (2004)
in {\em Proceedings of the SPIE}, {\em 5491}, 454.

\refs Clarke C. J., Gendrin A., and Sotomayor M. (2001) 
\mnras {\em 328}, 485-491.

\refs Crovisier J., Akeson R., Wizinowich P., Shao M., Acton S. 
et al. (1997) {\em Science}, {\em 275}, 1904-1907.

\refs D'Alessio P., Calvet N., Hartmann L., Muzerolle J., and Sitko M. (2004) in
{\em Star Formation at High Angular Resolution, IAU Symposium},
(M. G. Burton, R. Jayawardhana, and T. L. Bourke, eds.), {\em 221}, 403-410.

\refs Danchi W. C., Tuthill P. G., and Monnier J. D. (2001)
\apj, {\em 562}, 440-445.

\refs Dominik C., Dullemond C. P., Waters L. B. F. M., and Walch S.
(2003) \aap, {\em 398}, 607-619.

\refs Dullemond C. P., Dominik C., and Natta A. (2001) \apj, {\em 560}, 957-969.

\refs Dullemond C. P. and Dominik C. (2004) \aap, {\em 417}, 159-168.

\refs Drew J., Busfield G., Hoare M. G., Murdoch K. A., Nixon C. A.
et al. (1997) \mnras, {\em 286}, 538-548.

\refs Eisner J. A., Busfield G., Hoare M. G., Murdoch K. A., and Nixon C. A. 
(2003) \apj, {\em 588}, 360-372.

\refs Eisner J. A., Lane B. F., Hillenbrand L. A., Akeson R. L., and Sargent A. I.
(2004) \apj, {\em 613}, 1049-1071.

\refs Eisner J. A., Hillenbrand L. A., White R. J., Akeson R. L., and Sargent A. I.
(2005) \apj, {\em 623}, 952-966.

\refs Eisner J. A., Chiang, E. I., and Hillenbrand, L. A. (2006) \apj, {\em in press}.

\refs Feldt M., Pascucci I., Chesnau O., Apai D., Henning Th.
et al. (2006) in {\em The Power of Optical / IR Interferometry: Recent
Scientific Results and 2nd Generation VLTI Instrumentation},
(F. Paresce, A. Richichi, A. Chelli, and F. Delplancke, eds.), ESO
Astrophysics Symposia, Springer-Verlag, Garching.

\refs Fridlund M. C. (2003) in {\em Proceedings of the SPIE}, 5491, 227.

\refs Gail H.-P. (2004) \aap, {\em 413}, 571-591.

\refs Garcia P. J. V., Glindeman A., Henning T., and Malbet
F. (eds.) (2004) {\em The Very Large Telescope Interferometer --
Challenges for the Future}, Astrophysics and Space Science Volume 286,
Nos. 1-2, Kluwer Academic Publishers, Dordrecht.

\refs Gil C., Malbet F., Schoeller M., Chesnau O., Leinert Ch. et al. 
(2006) in {\em The Power of Optical / IR Interferometry: 
Recent Scientific Results and 2nd Generation VLTI Instrumentation},
(F. Paresce, A. Richichi, A. Chelli, and F. Delplancke, eds.), ESO
Astrophysics Symposia, Springer-Verlag, Garching.

\refs Hamilton C., Herbst W., Shih C., and Ferro A. J.
(2001) \apj, {\em 554}, L201-L204.

\refs Hanner M. S., Lynch D. K., and Russell R. W. (1994) \apj, {\em 425}, 274-285.

\refs Hale D., Bester M., Danchi W. C., Fitelson W., Hoss, S.
et al. (2000) \apj, {\em 537}, 998-1012.

\refs Hartmann L. and Kenyon S. J. (1985) \apj, {\em 299}, 462-478.

\refs Hartmann L. and Kenyon S. J. (1996) {\em Ann. Rev. Astron. Astrophys.}, {\em 34}, 207-240.

\refs Hartmann L., Calvet N., Gullbring E., and D'Alessio P. (1998) \apj, {\em 495}, 385.

\refs Herbig G. H., Petrov P. P., and Duemmler R. (2003) \apj, {\em 595}, 384-411.

\refs Hillenbrand L.~A., Strom S.~E., Vrba F.~J., and Keene J. (1992). 
\apj, {\em 397}, 613-643.

\refs Hillenbrand L. A. and White R. J. (2004) \apj, {\em 604}, 741-757.

\refs Hinz P. M., Hoffmann W. F., and Hora J. L. (2001) \apj, {\em 561}, L131-L134.

\refs Hollenbach D., Yorke H. W., and Johnstone D. (2000) 
in {\em Protostars and Planets IV}, (Mannings, V., Boss, A. P., and
Russell, S. S. eds.), pp. 401-428, Univ. of Arizona Press,
Tucson.

\refs Ireland M., Tuthill P. G., Davis J., and Tango W. (2005) \mnras, {\em 361}, 337-344.

\refs Isella A. and Natta A. (2005) \aap, {\em 438}, 899-907.

\refs Jayawardhana R., Fisher S., Hartmann L., Telesco C., Pina R. 
et al. (1998) \apj, {\em 503}, L79.

\refs Johns-Krull C. M. and Valenti V. A. (2001) \apj, {\em 561}, 1060-1073.

\refs Kemper F., Vriend W. J., and Tielens A. G. G. M. (2004) \apj, {\em 609}, 826-837.

\refs Kenyon S. J. and Hartmann L. W. (1987) \apj, {\em 323}, 714-733.

\refs Kenyon S. J. and Hartmann L. W. (1991) \apj, {\em 383}, 664-673.

\refs Koerner D. W., Ressler M. E., Werner M. W., and Backman D. E. (1998) 
\apj, {\em 503}, L83.

\refs Kuchner M. and Lecar M. (2002) \apj, {\em 574}, L87-L89.

\refs Lachaume R., Malbet F., and Monin J.-L. (2003) \aap, {\em 400}, 185-202.

\refs Lawson P. (ed.) (2000)
{\em Principles of Long Baseline Interferometry}, published by
National Aeronautics and Space Administration, Jet Propulsion
Laboratory, California Institute of Technology, Pasadena.

\refs Leinert Ch., Haas M., Abraham P., and Richichi A. (2001)
\aap, {\em 375}, 927.

%\refs Leinert Ch. (2004) in {\em Proceedings of the SPIE}, {\em 5491}, 19.

\refs Leinert Ch. Graser U., Przygodda F., Waters L. B. F. M., Perrin G. et al.
(2003) \apss, {\em 286}, 1, 73-83.

\refs Leinert Ch., van Boekel R., Waters L. B. F. M., Chesneau O., Malbet F. 
et al. (2004) \aap, {\em 423}, 537-548.

\refs Lin D. N. C., Bodenheimer P., and Richardson D. C. (1996) {\em Nature}, {\em 380}, 606-607.

\refs Liu W. M., Hinz P. M., Meyer M. R., Mamajek E. E. 
et al. (2003) \apj, {\em 598}, L111-L114.

\refs Liu W. M., Hinz, P. M., Hoffmann W. F., Brusa G., Miller D. 
et al. (2005) \apj, {\em 618}, L133-L136.

\refs Lynden-Bell D. and Pringle J. E. (1974) \mnras, {\em 168}, 603-637.

\refs Malbet F. and Bertout C. (1995) \aaps, {\em 113}, 369.

\refs Malbet F., ., Berger J.-P., Colavita M. M., Koresko C. D., Beichman C. 
et al. (1998) \apj, {\em 507}, L149-L152.

\refs Malbet F., Driebe T. M., Foy R., Fraix-Burnet D., Mathias P. 
et al. (2004) in {\em Proceedings of the SPIE}, {\em 5491}, 1722.

\refs Malbet F., Lachaume R., Berger J.-P., Colavita M. M., di Folco, E. 
et al. (2005) \aap, {\em 437}, 627-636.

\refs Malbet F., Benisty M., de Wit W. J., Kraus S., Meilland A.
et al. (2006) \aap, {\em in press}.

\refs Mathieu R. D. (1994) {\em Ann. Rev. Astron. Astrophys.}, {\em 32}, 465-530.

\refs McCabe C., Duchene G., and Ghez A. (2003) \apj, {\em 588}, 2, L113-L116.

\refs Meeus G., Waters L. B. F. M., Bouwman J., van den Ancker M. E., Waelkens C. 
et al. (2001) \aap, {\em 365}, 476-490.

\refs Millan-Gabet R., Schloerb F. P., Traub W. A., Malbet F., Berger 
J.-P., and Bregman, J. D. (1999) \apj, {\em 513}, L131-L143.

\refs Millan-Gabet R., Schloerb F. P., and Traub W. A. (2001)
\apj, {\em 546}, 358-381.

\refs Millan-Gabet R., Monnier J. D., Akeson R. L., Hartmann L., Berger J.-P.  
et al. (2006) \apj, {\em in press}.

\refs Monnier J. D. (2000) in {\em Principles of Long Baseline Interferometry}
(Lawson P. ed.), pp. 203-226, published by National Aeronautics and Space
Administration, Jet Propulsion Labratory, California Institute of
Technology, Pasadena.

\refs Monnier J. D. and Millan-Gabet R. (2002) \apj, {\em 579}, 694-698.

\refs Monnier J. D. (2003) in {\em Reports on Progress in Physics},
{\em Vol. 66}, Num. 5, 789-897

\refs Monnier J. D., Berger J.-P., Millan-Gabet R., and Ten Brummelaar T. A.
(2004a) in {\em Proceedings of the SPIE}, {\em 5491}, 1370.

\refs Monnier J. D., Tuthill P. G., Ireland M. J., Cohen R., and 
Tannirkulam A. (2004b), in {\em American Astronomical Society Meeting
2005}, 17.15.

\refs Monnier J. D., Millan-Gabet R., Billmeier R., Akeson R. L., Wallace D., Berger J.-P.
et al.  (2005a) \apj, {\em 624}, 832-840.

\refs Monnier J. D., Pedretti E., Millan-Gabet R., Berger J.-P., Traub, W. 
et al. (2005b) in {\em Protostars and Planets V Poster Proceedings}
{\tt http://www.lpi.usra.edu/meetings/ \\ppv2005/pdf/8238.pdf}

\refs Muzerolle J., Calvet N., Hartmann L., and D'Alessio P. (2003) \apj, {\em 597}, L149-L152

\refs Muzerolle J., D'Alessio P., Calvet N., and Hartmann L. (2004) \apj, {\em 617}, 406-417.

\refs Natta A., Grinin V., and Mannings V.  (2000)
in {\em Protostars and Planets IV}, (Mannings, V., Boss, A. P., and
Russell, S. S. eds.), pp. 559-588, Univ. of Arizona Press,
Tucson.

\refs Natta A., Prusti T., Neri R., Wooden D., Grinin V. P. 
et al. (2001) \aap, {\em 371}, 186-197.

\refs F. Paresce, A. Richichi, A. Chelli, and F. Delplancke (eds.)
(2006) {\em The Power of Optical / IR Interferometry: Recent
Scientific Results and 2nd Generation VLTI Instrumentation}, ESO
Astrophysics Symposia, Springer-Verlag, Garching.

\refs Perrin G. and Malbet F. (eds.) (2003)
{\em Observing with the VLTI}, EAS Publications Series, {\em Volume
6}.

\refs Quanz S. P. (2006) in {\em The Power of Optical / IR Interferometry: 
Recent Scientific Results and 2nd Generation VLTI Instrumentation},
(F. Paresce, A. Richichi, A. Chelli, and F. Delplancke, eds.), ESO
Astrophysics Symposia, Springer-Verlag, Garching.

\refs Quirrenbach A. (2001) {\em Ann. Rev. Astron. Astrophys.}, {\em 39}, 353-401.

\refs Quirrenbach A., Albrecht S., and Tubbs R. N. (2006)
in {\em Stars with the B[e] phenomenon}, (Michaela Kraus and Anatoly
S. Miroshnichenko, eds.), ASP Conference Series, {\em in press}.

\refs Rice W. K. M., Wood K., Armitage P. J., Whitney B. A., and Bjorkman J. E.
(2003) \mnras, {\em 342}, 79-85.

\refs Serabyn E., Colavita M. M., and Beichman C. A.
(2000) in {\em Thermal Emission Spectroscopy and Analysis of Dust,
Disks, and Regoliths} (Michael L. Sitko, Ann L. Sprague, and David
K. Lynch, eds.), {\em 196}, pp. 357-365, ASP Conf. Series, San
Francisco.

\refs Serabyn E., Booth A. J., Colavita M. M., Creech-Eakman M. J., Crawford S. L.
et al. (2004) in {\em Proceedings of the SPIE}, {\em 5491}, 806.

%\refs Schaefer G. H., Simon M., and Prato L. (2005) in 
% {\em The Power of Optical / IR Interferometry: Recent Scientific
%Results and 2nd Generation VLTI Instrumentation}, (F. Paresce,
%A. Richichi, A. Chelli, and F. Delplancke, eds.), ESO Astrophysics
%Symposia, Springer-Verlag, Garching.

\refs Shu F., Najita J., Ostriker E., Wilkin F., Ruden S.
et al. (1994) \apj, {\em 429}, 781-807.

\refs Simon M., Dutrey A., and Guilloteau S. (2000) \apj, {\em 545}, 1034-1043.

\refs Stapelfeldt K., Watson A. M., Krist J. E., Burrows C. J., Crisp D. 
et al. (1999) \apj, {\em 516}, 2, L95-L98.

\refs ten Brummelaar T. A., McAlister H. A., Ridgway S. T., Bagnuolo W. G. Jr., Turner N. H. 
et al. (2005) \apj, {\em 628}, 453-465.

\refs Tokovinin, A. (1999) {\em Astron. Lett.}, {\em 25}, 669-671.

\refs Traub W. A., Berger. J.-P., Brewer M. K., Carleton N. P., Kern P. Y.
et al. (2004) in {\em Proceedings of the SPIE}, {\em 5491}, 482.

\refs Tuthill P., Monnier J. D., and Danchi, W. C. (2001)
{\em Nature}, {\em 409}, 1012-1014.

\refs Tuthill P., Monnier J. D., Danchi W. C., Hale D. D. S., and Townes C. H.
(2002) \apj, {\em 577}, 826-838.

\refs van Boekel R., Min M., Leinert Ch., Waters L. B. F. M., Richichi A. 
et al. (2004) {\em Nature}, {\em 432}, 479-482.

\refs van Boekel R., Dullemond C. P., and Domink C. (2005) \aap, {\em 441}, 563-571.

\refs Vinkovi\'{c} D., Ivezi\'{c} \v{Z}., Jurki\'{c} T., and Elitzur M. (2006)
\apj, {\em 636}, 348-361.

\refs Wilkin F. P. and Akeson R. L. (2003) \apss, {\em 286}, 145-150.

\refs Wootten A. (2003) in {\em Proceedings of the SPIE}, {\em 4837}, 110.

%\clearpage

%%%%%%%%%%%%%%%%%%%%%%%%%%%%%%%%%%%%%%%%%%%%%%%%%%%%%%%%%%%%%%%%%%%%%%%

%\input{ysos_table}

%%%%%%%%%%%%%%%%%%%%%%%%%%%%%%%%%%%%%%%%%%%%%%%%%%%%%%%%%%%%%%%%%%%%%%%

\end{document}